\documentclass[aps,amsmath,amssymb,showpacs,showkeys]{revtex4} 
\usepackage[dvips]{graphicx,color}%
\usepackage{times}
\usepackage{multirow}
\usepackage{xcolor}
\usepackage{color,soul}
\usepackage[
  colorlinks=true,
  urlcolor=magenta,
  linkcolor=red,
  citecolor=blue]{hyperref}
\newcommand{\orcid}[1]{\href{https://orcid.org/#1}{\includegraphics[width=8pt]{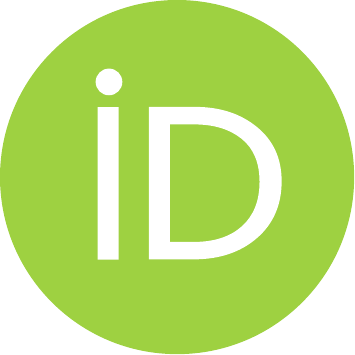}}}

\begin{document}
\title{Radial oscillations and gravitational wave echoes of strange stars with nonvanishing lambda}

\author{Jyatsnasree Bora \orcid{0000-0001-9751-5614}}
\email[Email: ]{jyatnasree.borah@gmail.com} 

\author{Umananda Dev Goswami \orcid{0000-0003-0012-7549}}
\email[Email: ]{umananda2@gmail.com}

\affiliation{Department of Physics, Dibrugarh University, Dibrugarh 786004, 
Assam, India}

\begin{abstract}
We study the effect of the cosmological constant on radial oscillations 
and gravitational wave echoes (GWEs) of non-rotating strange stars. 
To depict strange star configurations we used two forms of equations of 
state (EoSs), viz., the MIT Bag model EoS and the linear EoS. By taking a 
range of positive and negative values of cosmological constant, the 
corresponding mass-radius relationships for these stars have been calculated. 
For this purpose, first we solved the Tolman-Oppenheimer-Volkoff (TOV) 
equations with a non-zero cosmological constant and then we solved the 
pressure and radial perturbation equations arising due to radial oscillations. 
The eigenfrequencies of the fundamental $f$-mode and first 22 pressure 
$p$-modes are calculated for each of these EoSs. Again considering the remnant 
of the GW170817 event as a strange star, the echo frequencies emitted by 
such stars in presence of the cosmological constant are computed. From these 
numerical calculations, we have inferred relations between cosmological 
constant and mode frequency, structural parameters, GWE frequencies of strange 
stars. Our results show that for strange stars, the effective range of 
cosmological constant is $10^{-15}\, \mbox{cm}^{-2} \le\,\Lambda\, \le 3\times 
10^{-13}\, \mbox{cm}^{-2}$.
\end{abstract}

\pacs{04.40.Dg,97.10.Sj}
\keywords{dense matter -- asteroseismology -- cosmological constant -- gravitational waves}

\maketitle

\section{Introduction}\label{intro} 
In 1916 Albert Einstein introduced the term cosmological constant $\Lambda$ 
as a modification in his field equation to achieve a static, stationary 
universe, which was believed to be the state of the universe at that time. 
This new term introduced by him is a dimensionful free parameter 
\cite{carroll}. But after the discovery of the redshift of stars and 
consequently an expanding universe by Hubble this idea of cosmological constant was quickly 
abandoned. Later, it was reintroduced to overcome the age crisis problem and to 
construct a universe satisfying the ``perfect cosmological principle" 
\cite{bondi, hoyle}. On the contrary, current astrophysical and cosmological 
observational data reveal that we are in the era of cosmological expansion and 
the expansion rate of the universe is increasing rather than decreasing 
\cite{riess, perlmutter}. One of the strong beliefs is that the present 
accelerated expansion of the universe is due to an exotic form of energy 
known as  dark energy. Although it suffers from the fine-tuning 
problem, $\Lambda$ has been considered as one of the prime 
candidates of dark energy in the sense that it forms the vacuum 
energy \cite{Peebles}. As dark energy pervades throughout the 
universe, the energy density inside the compact objects should be affected by 
it, and hence it is necessary to see the effect of dark energy content on the 
behaviour of such objects. So, if we consider the contribution of cosmological 
constant $\Lambda$ in compact objects it will be related to the acceleration of 
the currently observable universe or dark energy. This possibility of 
non-zero $\Lambda$ and it, as a dominating energy density of the universe, is a 
fascinating problem in nowadays research. 

One of the most intriguing parts of the issue of compact stars is the study of 
unique strange stars. Since the last decade, the subject of strange stars has 
attracted much attention from the astrophysicist community. 
These hypothetical stars are unique in the sense that the structural behaviour 
of such objects rarely matches with that of other compact objects, like 
white dwarfs and neutron stars. Due to their unique structural behaviour, the 
strange stars can be regarded as excellent natural laboratories to study, test, 
or perhaps constrain different modified theories of gravity under 
extreme conditions that cannot be reached from the earth-based 
laboratories. By hypothesis, strange matters are the true ground state of the 
hadrons and hence they are stable matters \cite{Witten1984}. 
So, it could explain the origin of the huge amount of energy released in 
superluminous supernovae (100 times brighter than normal supernovae) 
\cite{ofek, ouyed}. Such strange matters are composed of deconfined quark 
matters, mainly of $u$, $d$, $s$ quarks and a small amount of electrons to 
maintain charge neutrality \cite{aclock, weber2005}.

Our present study consists of two parts. In the first part, we are seeking 
correlations between the cosmological constant $\Lambda$ and radial 
oscillation frequencies of strange stars. From long ago, the light 
variation in a pulsating star has been used to investigate the physical 
properties like mass, radius of a star. This indirect approach to understanding 
the physical properties of a star is known as asteroseismology 
\cite{handler2012}. From the birth to the end of a star, 
nearly every star undergoes some kind of pulsation. Such asteroseismic 
behavioural studies of stars are mainly of two types: radial and non-radial. 
For the case of compact objects like white dwarfs, neutron 
stars, and strange stars, such radial asteroseismic behaviours 
are reported earlier in \cite{Chandrasekhar1964a, Chandrasekhar1964b, 
Chanmugam1977, datta1992}. Here, we stick to our study in the case of radial 
oscillations only of strange stars.

In the second part of this study, we investigate the effect of $\Lambda$ on the 
gravitational wave echoes (GWEs) of strange stars. It has been recently 
suggested that some ultra-compact post-merger objects can emit GWEs \cite{pani, 
manarelli, abedi, urbano}. Since strange stars are known to be very compact, we 
describe strange stars as ultra-compact objects by using stiffer equations of 
state (EoSs). In the recent detection of gravitational waves (GWs) from the 
binary neutron star merging event GW170817, the nature of the final massive 
remnant formed is still not confirmed. We consider it as a strange star and 
evaluate the corresponding echo frequencies emitted by such stars in presence 
of the cosmological constant $\Lambda$. The ultra-compact stars are those, 
whose compactness, $\mbox{C}=\mbox{M}/\mbox{R}$ exceeds $1/3$ \cite{urbano}. 
Again to be an eligible candidate to emit echo frequencies, the star must 
possess a photon sphere. This is the region above the surface of the 
star at a distance of $\mbox{R}=3\,\mbox{M}$ \cite{pani}. However, the 
compactness should not exceed Buchdahl's limit $\mbox{R}=9/4\,\mbox{M}$. 
Buchdahl's limit is the upper limit of compactness of fluid 
stars which describes the maximum amount of mass that can exist in a sphere 
before it must undergo the gravitational collapse \cite{urbano}.

In presence of cosmological constant, compact stars are studied earlier in 
\cite{nayak,Bordbar,Largani,Liu,arbanil}. In the article \cite{bohmer}, the 
equations describing small radial oscillations of relativistic stars in 
presence of a cosmological constant $\Lambda$ were reported. They have also 
studied the impact of cosmological constant $\Lambda$ on the critical adiabatic 
index. Again Hossein et al. \cite{Sk} studied the anisotropic compact stars by 
considering a variable cosmological constant with a static spherically 
symmetric spacetime described by Krori-Barua metric \cite{kb}. A similar study 
was done by Kalam et al. for the anisotropic compact stars with the de Sitter 
spacetime \cite{kalam}.  In 2014, O. Zubairi and F. Weber derived the modified 
stellar structure equations to account for a finite value of the cosmological 
constant in spherically symmetric mass distributions \cite{zubairic1}. In 
\cite{zubairic}, such equations are derived for the compact stellar structures 
such as quark star and neutron stars in presence of the cosmological constant 
$\Lambda$, and the role of the cosmological constant on stars mass, 
radius, pressure, and density profiles along with the gravitational 
redshifts were discussed. For compact stars with an equation of state provided 
by the quark-meson coupling (QMC) model, such effects of the cosmological 
constant are discussed in \cite{nayak}. Recently, for stable relativistic 
polytropic objects, the effect of the cosmological constant is reported in 
\cite{arbanil}. A more detailed analysis of the stability of polytropic spheres 
in the presence of a cosmological constant can be found in \cite{posada}. For 
the general relativistic case with a vanishing cosmological constant, the 
radial oscillation modes and echo frequencies for strange stars are studied in 
\cite{jb}. The present study is the extension of our previous study \cite{jb} 
in which we had taken into account the dependence of radial oscillation 
frequencies and GWE frequencies on different model parameters. This work will 
focus on the impact of the cosmological constant on radial oscillation 
modes and echo frequencies of strange stars.

Motivated from the previous works mentioned above, we aim at exploring the 
effect of the cosmological constant on strange stars' structural 
behaviour, radial oscillation, and echo frequencies emitted by them and 
also to find boundary values of it. Here, we take the strange star as a probe 
to explore the consequences of a non-vanishing cosmological constant and allow 
it to vary inside the star. Also, the cosmological constant is taken as a 
matter distribution source of the star and we show that the compactness of the 
star varies with the cosmological constant. In turn, it implies that the 
cosmological constant varies inside such dense stars depending on their 
compactness. The presence of cosmological constant $\Lambda$ describes well 
the strange star configurations. In particular, we study 
the structural change in strange stars due to a non-vanishing cosmological 
constant, focusing on its effects over the radial oscillations and GWEs of 
strange stars. Here, we have solved the Einstein field equation for some finite 
values of cosmological constant $\Lambda$ for spherically symmetric 
mass distribution. By solving the Tolman-Oppenheimer-Volkoff (TOV) equations we 
have obtained the mass-radius relationships of strange stars in presence of a 
non-vanishing cosmological constant. Then we have computed the fundamental $f$-
mode and first 22 pressure $p$-mode of radial oscillations, and GWE frequencies 
emitted by strange stars in two types of EoSs.

We have organized the rest of the paper as follows. In Sec.\ \ref{gr} the 
general relativistic formulations including TOV equations and perturbation 
equations are discussed briefly. In Sec.\ \ref{eos} the considered EoSs are 
described along with a brief note on the cosmological constant. The emission of 
GWEs from an ultra-compact object is discussed in Sec.\ \ref{echo}. 
In Sec.\ \ref{numerical} we have discussed the results that are obtained from 
this study, which is finally followed by the concluding section, 
i.e.\ Sec.\ \ref{conclusion}. Here we follow the natural unit system by 
considering $c=\hbar=1$ and $G=1$ with the metric convention $(-,\,+,\,+,\,+)$.

\section{General relativistic formulation}\label{gr}
\subsection{Equations for stellar structure}\label{tov}
In this subsection, we discuss briefly about the strange stars in general 
relativity (GR) in a spacetime with a cosmological constant, i.e.\ with
 $\Lambda \neq 0$. The Einstein's field equation with the cosmological 
constant reads,
\begin{equation}
	\label{eq1}
	G_{\mu\nu}+\Lambda g_{\mu\nu}=8\pi T_{\mu\nu}.
	\end{equation}
We consider the strange star as isotropic, stable, and 
non-rotating mass distribution and the perfect fluid inside the strange star 
is described by the stress-energy tensor,
\begin{equation}
	\label{eq2}
	T_{\mu \nu} = (p+\rho)U_{\mu}U_{\nu}+p\,g_{\mu \nu},
	\end{equation}
where $p$ is the fluid pressure, $\rho$ is the fluid energy density and 
$U_{\mu}$ are its four-velocities. The line element for a 
spherically symmetric mass distribution has the form:
\begin{equation}
	\label{eq3}
	ds^{2} = -\,e^{\chi(r)}dt^{2}+e^{\lambda(r)}dr^{2}+r^{2}\,d\theta^{2}+r^{2}
	\sin^{2}\theta\,d\phi^{2}
    \end{equation}
with the unknown metric functions $\chi(r)$ and $\lambda(r)$ being dependent 
on the radial coordinate $r$ only. For the exterior problem the field equation
\eqref{eq1} gives,
\begin{equation} 
	\label{eq4}
	e^{\chi(r)}=e^{-\lambda(r)}=1-\dfrac{2M}{r}-\dfrac{\Lambda r^{2}}{3},
	\end{equation}
where $M$ is the total mass of the star. Similarly for the interior problem we 
have,
\begin{equation}
	\label{eq5}
	e^{\chi(r)}=e^{-\lambda(r)}=1-\dfrac{2m(r)}{r}-\dfrac{\Lambda r^{2}}{3}.
	\end{equation}
Here $m (r)$ is the mass function within the radius $r$. On matching the 
exterior and interior solutions at the surface of the star we get, $m(R)=M$, 
the total mass of the stellar configuration with $R$ being the radius of the 
star.

The addition of cosmological constant $\Lambda$ in Einstein's field equation 
(\ref{eq1}) will change the structure of compact objects. Now solving  
Einstein's field equation for the energy-momentum tensor (\ref{eq2}) we get 
the stellar structure equations, known as the TOV equations \cite{tolman, tov} 
with the introduction of cosmological constant \cite{zubairic} as
\begin{equation}
	\label{eq6}
	\dfrac{dm}{dr} = 4\pi \rho (r) r^{2},
	\end{equation}
\begin{equation}
	\label{eq7}
	\dfrac{dp}{dr} = -(\rho + p) \dfrac{m + 4 \pi p r^{3}-\dfrac{\Lambda r^{3}}{3}}
	{r^{2}\left(1-\dfrac{2m(r)}{r} -\dfrac{\Lambda r^{2}}{3}\right)},
	\end{equation}
\begin{equation}
	\label{eq8}
	\dfrac{d\chi}{dr} = - \dfrac{2}{\rho + p} \dfrac{dp}{dr}.
	\end{equation}
To find static equilibrium stellar configurations equations (\ref{eq6}) - 
(\ref{eq8}) are to be integrated along the radial coordinate $r$ with the 
initial conditions: $m(r=0)=0$, $p(r=0)=p_{c}$ and $\rho(r=0)=\rho_{c}$. The 
radius of the star can be determined by using the fact that the pressure 
vanishes at the surface of the star, i.e.\ $p(r=R)=0$ (see Fig.\ \ref{fig10}). 
These equations for stellar structure can be solved for a given EoS. Solutions 
of these equations can lead us to know about the stellar mass, radius, 
pressure, density profile, and the gravitational redshift.

\subsection{Radial stability equations}\label{radial}
The theory of infinitesimal, adiabatic, radial oscillations and the radial 
stability of relativistic stars was first derived by S.\ Chandrasekhar in 
1964 \citep{Chandrasekhar1964a, Chandrasekhar1964b}. The solutions of these 
perturbation equations give information about the eigenfrequencies of 
radial oscillations. In the presence of cosmological constant $\Lambda$, 
the radial instability equations were investigated earlier 
in \cite{bohmer,stuchlik}. By introducing two dimensionless parameters 
$\xi= \Delta r/r$ and $\eta = \Delta p/p$, where $\Delta r$ 
is the radial perturbation and $\Delta p$ is the corresponding Lagrangian 
perturbations of the pressure, Bh\"omer and Harko \cite{bohmer} 
presented the radial and pressure perturbation equations as
\begin{equation}
	\label{eq9}
	\small {\dfrac{d\xi}{dr} = -\, \dfrac{1}{r}\left(3\xi + \dfrac{\eta}{\gamma}
	\right) - \dfrac{dp}{dr}\dfrac{\xi}{p+\rho}},
	\end{equation}
	\begin{align}
	\label{eq10}
	\dfrac{d\eta}{dr} =  \xi\left[\omega^{2}e^{\lambda-\chi}(1+\dfrac{\rho}{p})r
	-\dfrac{4}{p}\dfrac{dp}{dr}+\;\dfrac{r}{p\,(\rho+p)}{\left(\dfrac{dp}{dr}
	\right)}^{2}-e^{\lambda}\left(8\pi-\dfrac{\Lambda}{p}\right)(\rho+p)\,r
	\right]\notag \\+\,\eta\left[\dfrac{1}{(\rho+p)}\dfrac{dp}{dr}-4\pi(\rho+p)
	\,r\,e^{\lambda}\right],
	\end{align}
where $\gamma=\dfrac{dp}{d\rho}(1+\rho/p)$ is the relativistic adiabatic 
index.

To study the stability of the stellar object against a small radial 
perturbation we have to integrate equations (\ref{eq9}) and (\ref{eq10}) from 
the centre to the boundary along with two boundary conditions: one at the 
centre and other at the surface of the star. As $r\rightarrow 0$, the 
coefficient of $1/r$ in equation (\ref{eq9}) must vanish to avoid the
singularity in the equation. So at the centre of the star we get the boundary 
condition as
\begin{equation}
\label{a}
3\gamma \xi + \eta = 0.
\end{equation} 
From this condition it can be inferred that $\Delta p = -\,3\gamma p \xi$ at 
the centre of the star, i.e.\ $(\Delta p)_{r\,=\,0} = - 
{(3\gamma p \xi)}_{r\,=\,0}$. Again when $r\rightarrow R$, i.e.\ at the 
surface of the star, $p\rightarrow 0$. Thus again to avoid singularity this 
will eventually require that the coefficient of $1/p$ in the equation 
(\ref{eq10}) must vanish at the boundary of the star. Using equation 
(\ref{eq7}) in equation (\ref{eq10}), it can be found from this requirement that
\begin{align}
\label{b}
\xi\left[\omega^{2}R\left(\dfrac{M}{R^2}-\dfrac{\Lambda R}{3}\right)^{-1}\left(1-\dfrac{2M}{R}-\dfrac{\Lambda R^2}{3}\right)^{-1}+R\left(\dfrac{M}{R^2}-\dfrac{\Lambda R}{3}\right)\left(1-\dfrac{2M}{R}-\dfrac{\Lambda R^2}{3}\right)^{-1}\right] \notag \\ + \xi\left[\Lambda R\left(\dfrac{M}{R^2}-\dfrac{\Lambda R}{3}\right)^{-1}+4\right]=0.
\end{align} 
This boundary condition implies that $\Delta p=\,0$ at the 
stellar surface, i.e.\ $(\Delta p)_{r\,=\,R} = 0$. These coupled differential equations constitute the Sturm-
Liouville type eigenvalue problem. These equations (\ref{eq9}) and (\ref{eq10}) 
together with the boundary conditions (equations (\ref{a}), (\ref{b})) determine the eigenvalues or eigenfrequencies $\omega$ of radial 
oscillations \cite{jb}. These equations are solved by using the shooting method 
as described in \cite{panotopoulus}. If $\omega$ is real, i.e.\ $\omega^{2} >0$, 
the configuration is stable and for $ \omega^{2} < 0$, the configuration becomes 
unstable against radial oscillations. Again for the stable fundamental mode ($f$-mode) i.e.\ for $\omega_{0}^{2} >0$, all other higher-order modes will also be 
stable.

To find the radial oscillation frequencies we first calculated the quantity $
\bar{\omega}=\omega\, t_{0}$, which is dimensionless, and here 
$t_{0}=1\,\mbox{ms}$. The oscillation frequencies are then calculated by 
\begin{equation}\label{eq16}
	\nu=\dfrac{\overline{\omega}}{2\pi}\;\;\mbox{kHz}.
	\end{equation}
	The obtained frequency $\nu$ is now allowed to take some eigenvalues $\nu_{n}$ 
	and for each values of $\nu_{n}$ we have obtained a specific oscillation mode of 
	the star \cite{jb}.

\section{Equations of State and the cosmological constant}\label{eos}
Before proceeding to discuss the role of the cosmological constant on strange 
stars' oscillation and echo frequencies emitted by them, we wish to add a few 
comments on our choice of stellar models. It is well known that the 
macroscopic properties of compact objects, such as mass and radius, depend 
crucially on the EoSs of ultra-compact matter, which is unfortunately not 
confirmed clearly to date. In this study, the chosen EoSs are those that (i) 
should be stiff enough to emit GWE, (ii) the model parameters associated with 
each EoS lie inside the desired range, and (iii) the mass-radius relations of 
which are within the accepted limits. To describe the stellar structure, in 
this present work, we are using two EoSs, viz., the MIT Bag model EoS and the 
linear EoS as mentioned above. 

The MIT Bag model corresponds to a relativistic gas of deconfined quark matter 
with energy density. This EoS satisfies all necessary criteria for strange 
matter while retaining an elegant simplicity \cite{haensel}. It was first 
used by Witten in 1984 \cite{Witten1984}. The form of this equation is 
\begin{equation}
\label{eq11}
p=\dfrac{1}{3}(\rho-4\,B). 
\end{equation}
In this EoS, the density at the stellar surface is given by 
$\rho(R)=4\,\mbox{B}$, where $p$ is the isotropic pressure, $\rho$ is the 
energy density and $B$ is the Bag constant. Instead of taking this usual 
form (\ref{eq11}) of the MIT Bag model EoS we have chosen the stiffer form 
of this EoS as

\begin{equation}
\label{eq12}
p=\rho-4\,B 
\end{equation}
The reason behind doing this is that in EoS of the form (\ref{eq11}) the 
compactness of the stellar structure obtained is not enough to emit GWE 
frequencies. Making it a stiffer one of the form \eqref{eq12} allows the star 
to get enough compactness to emit GWE frequencies. As reported in 
\cite{aziz}, the acceptable range of Bag constant is 
$(133.68\,\mbox{MeV})^4<\mbox{B}<(222.54\,\mbox{MeV})^4$. 
So in this work we choose $\mbox{B}$ as $(190\,\mbox{MeV})^4$, which is well 
inside the desired range.

Other EoS we have used is the linear EoS of the form:
\begin{equation}
\label{eq13}
p=b\,(\rho-\rho_{s}),
\end{equation}
where $b$ is the linear constant and $\rho_{s}$ is the surface energy density 
\cite{rosinska}. This EoS was developed by Dey et al.\ in 1998 \cite{dey}. We 
have chosen the linear constant $b=0.910$ and the corresponding value of the 
surface energy density $\rho_{s}$ is taken. While choosing the constant 
value, we have kept in mind the conditions for echoing GWs, which restrict 
that the compactness should be larger than $1/3$, and also it should respect 
the causality condition ($b\leq 1$) \cite{jb}. Our chosen value of $b$ lies 
under these two restrictions.

The dimensionful parameter $\Lambda$ introduced by Einstein in his field 
equation has the unit of $\mbox{(length)}^{-2}$. We already have mentioned in 
Sec.\ \ref{intro} about the various studies on compact objects in presence 
of cosmological constants \cite{nayak,Bordbar,Largani,Liu,arbanil}. 
Specifically, we would like to mention here that in Ref. \cite{zubairic1}, O.\ 
Zubairi and F.\ Weber obtained that the properties of compact objects in our 
Universe, where $\Lambda$ is very small (predicted via cosmological 
observations), do not depend on the cosmological constant. They observed that 
to yield an observable effect, $\Lambda$ would have to be of the order of 
nuclear density scales $\sim 10^{-14}\, \mbox{cm}^{-2}$. For neutron stars, a 
maximum mass of $\sim $ 1.68 $\mbox{M}_{\odot}$ was reported by G.H. Bordbar et 
al.\ \cite{Bordbar} using $\Lambda\sim 10^{-18}\, \mbox{cm}^{-2}$. For the case 
of white dwarfs, a recent study demands the upper limit on $\Lambda$ as, 
$\Lambda < 3\times10^{-18}\, \mbox{cm}^{-2}$ \cite{Liu}. Largani et al.\ found 
that for neutron stars to have an observable effect of $\Lambda$ in their 
structure, the typical values of $\Lambda$ to be $\sim 10^{-12}\, 
\mbox{cm}^{-2}$\cite{Largani}. In the article \cite{arbanil}, the authors 
reported the stellar configuration of polytropic spheres using $\Lambda$ values 
within the range of $-10^{-15}\leq\Lambda\leq10^{-15}\, \mbox{cm}^{-2}$.
In \cite{bohmer}, for a star composed of matter having a density equivalent 
to nuclear energy density, the upper limit of $\Lambda$ was reported to be 
less than $3\times 10^{-13}\, \mbox{cm}^{-2}$. Thus, motivated by the search of 
new equilibrium configurations, corresponding radial oscillations, and also the 
properties of GWE frequencies of strange stars, in this study we have 
considered a cosmological constant value larger than the one predicted by 
cosmological observations \cite{planck}. We have chosen a set of positive and 
negative values of $\Lambda$ to see its impact on radial oscillation modes 
and echo frequencies of strange stars. Our chosen values of $\Lambda$ are in 
multiples of the $\epsilon$ as $-\, 150\,\epsilon$, $-\,100\,\epsilon$, $-\, 50\,\epsilon$, $5\,\epsilon$, $10\, \epsilon$, and $15\,\epsilon$. These values 
are in the units of nuclear energy density scale, $\epsilon = 140\, \mbox{MeV/\,fm}^{3}$. The study with positive values of cosmological constant $\Lambda$ 
(which corresponds to de Sitter space) gives a more realistic description of 
strange stars. These are relevant from the observational point of view and also 
used in the context of the dark energy model of the universe. Again the 
negative cosmological constant $\Lambda$ corresponding to anti-de Sitter space 
is important for anti-de Sitter/conformal field theory correspondence. So this 
work is a generalized work that is going to contribute significantly in both 
observational regime and conformal field theory regime. Again, it is worthy to 
mention that besides the case of the configuration of compact stellar objects, 
large values of the cosmological constant have been considered in other types 
of astrophysical situations also \cite{arbanil}. For instance, in the case of 
accretion in primordial black holes during the very early universe, the 
cosmological constant can take values many orders of magnitude greater than 
that predicted by current observational data \cite{zstuch}. As described by 
Carroll in classical GR, there is no preferred choice for the length scale 
defined by $\Lambda$ \cite{carroll}. Indeed, the cosmological constant is a 
measure of the energy density of the vacuum or the state of lowest energy. This 
value cannot be calculated with any confidence. So, it brings flexibility in 
choosing scales of various contributions to the cosmological constant \cite{carroll}.

\section{Gravitational wave echoes}\label{echo}
As mentioned in Sec.\ \ref{intro}, an important property of ultra-compact 
objects is that some of them can emit GWE frequency. It is a very interesting 
property of such stars. This happens due to the presence of a photon sphere 
around the surface of the star. More precisely, GWEs originate from the GWs 
that are trapped between the photon sphere and the surface of the ultra-compact star \cite{abedi}. As mentioned in \cite{abedi}, such type of echoes 
has two natural frequencies: the harmonic or resonance frequencies and the 
black hole ringdown or quasi-normal mode (QNM) frequencies.

In this study, we have chosen the final remnant of the GW170817 event as a 
strange star and calculated the GWEs emitted by such ultra-compact objects. The 
GWE frequency can be approximately estimated from the inverse of the time taken
by a massless test particle to travel from the unstable light ring to the 
centre of the star. This time is the characteristic echo time and can be 
expressed as \cite{urbano}
\begin{equation}
	\label{eq19}
	\tau_{echo}\equiv\int_0^{3M}\!\!\!\! e^{\,(\lambda(r)-\chi(r))/2}
	\; \mathrm{d}r.
	\end{equation} 
Using the relation for $e^{\,-\lambda(r)}$ from equation (\ref{eq5}) in 
presence of a non-zero cosmological constant $\Lambda$, we can obtain the 
following expression for characteristic echo time,
\begin{equation}
	\label{eq20}
	\tau_{echo}=\int_0^{3M}\!\!\!\! \dfrac{1}{\sqrt{e^{\,\chi(r)}\left(1-
	\dfrac{2m(r)}{r}-\dfrac{\Lambda r^2}{3}\right)}}
	\; \mathrm{d}r.
	\end{equation} 
In this equation, the term $m(r)$ and $\chi(r)$ can be obtained from the 
solution of TOV equations with non-vanishing cosmological constant 
$\Lambda$, i.e.\ of equations (\ref{eq6}) - (\ref{eq8}). Finally, the 
characteristic echo frequency can be calculated by using the relation, 
$\omega_{echo} \approx \pi/\tau_{echo}$.


\section{Numerical results}\label{numerical}
\begin{figure*}
	\centerline{
	\includegraphics[scale=0.32]{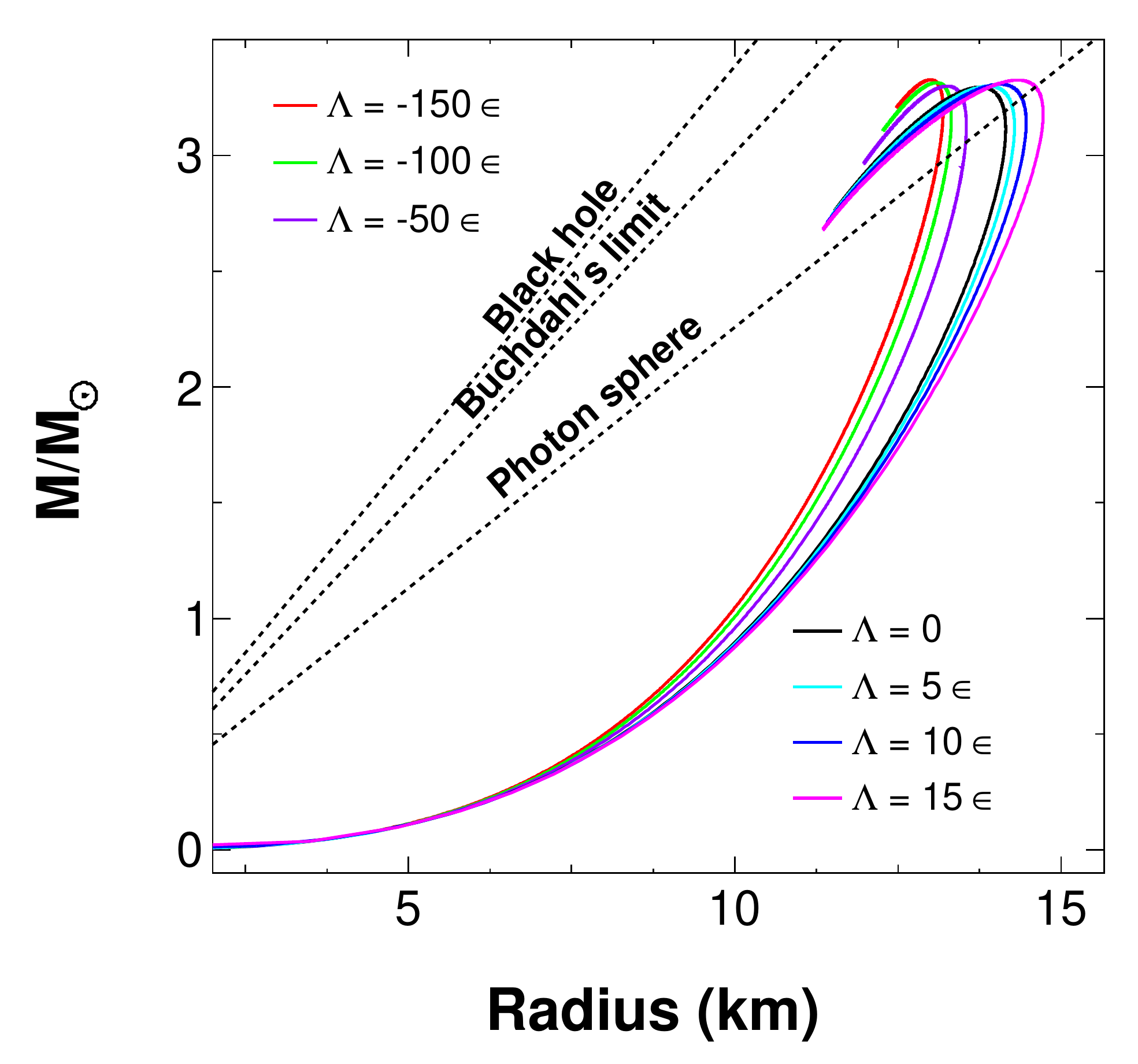}\hspace{0.5cm}
	\includegraphics[scale=0.32]{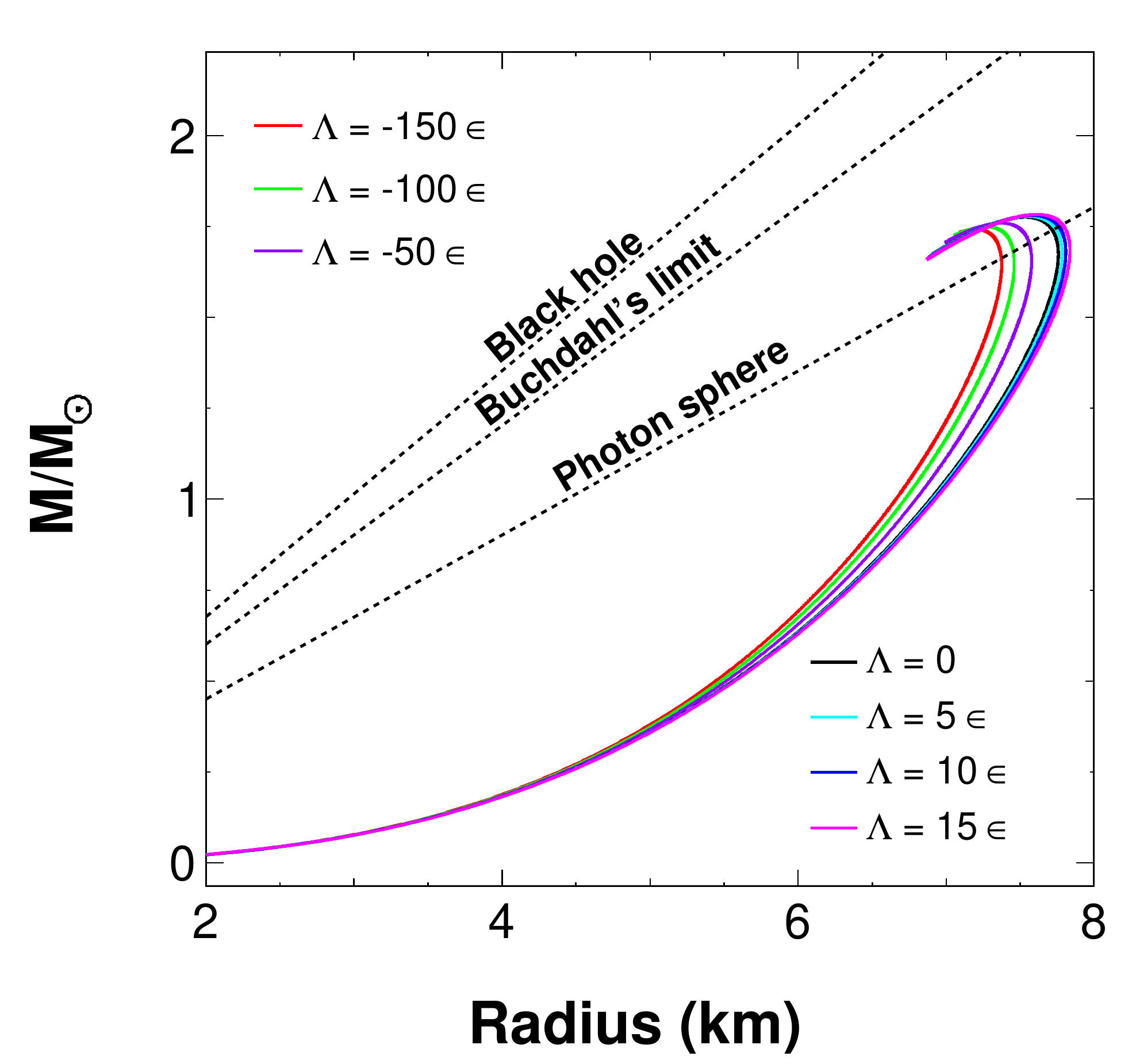}}	
	\vspace{-0.2cm}
	\caption{Mass-radius relationships of strange stars for the MIT Bag 
model EoS (left plot) and linear EoS (right plot) with different cosmological 
constants showing photon sphere limit, Buchdahl's limit and black hole limit 
lines for negative, zero and positive $\Lambda$ values. Mass of stars is 
expressed in terms of solar mass $\mbox{M}_{\odot}$.} 
	\label{fig1} 
	\end{figure*}
As mentioned earlier, in comparison to other compact objects the structural 
properties of strange stars are unique. The mass-radius relationship of these
stars follows a definite pattern than that of neutron stars. Strange matter 
can exist in lumps with the size of few fermis to the size of $\sim 10$ km 
radius strange stars \cite{aclock}. Using the 
stiffer MIT Bag model EoS and linear EoS we have plotted the sequences of 
mass-radius relationships for such stellar configurations. In this study, with 
the MIT bag model EoS, the maximum mass is obtained as 
$\approx 3.33\,\mbox{M}_{\odot}$ and the corresponding maximum radius is 
$\approx 14.34\,\mbox{km}$ and this corresponds to the $\Lambda$ value $15\,\epsilon$. The maximum mass obtained for the linear EoS is $\approx 1.78\,\mbox{M}_{\odot}$ and the maximum radius is $\approx 7.61\,\mbox{km}$ 
corresponding to the same $\Lambda$ value. The first plot of Fig.\ \ref{fig1} 
shows the relationship for the MIT Bag model EoS and for the linear EoS it is 
shown in the second plot. As shown in these two plots, for much of these 
sequences strange stars are following the $\mbox{M}\propto \mbox{R}^{3}$ 
relation. However, for neutron stars, radii decrease with increasing mass for 
much of their ranges. For both of the models, with decreasing $\Lambda$, 
stiffer configurations are obtained. The stiffness is observed to be maximum 
with $\Lambda=-150\,\epsilon$ and for the MIT Bag model it is higher than the
linear EoS. Linear EoSs are giving smaller stellar configurations than the MIT 
Bag model EoSs. In case, if we choose the general form of the MIT Bag model 
EoS, then the compactness will not be sufficient to overcome the photon sphere 
limit as pointed out earlier. The stiffer form of the MIT Bag model and linear 
EoSs are giving compact enough configurations to cross the photon sphere limit 
but not Buchdahl's limit. From Fig.\ \ref{fig1}, it is clear that 
the sequences of stellar configurations that are obtained can 
emit GWE frequencies with the considered $\Lambda$ values. 

\begin{figure*}
        \centerline{
        \includegraphics[scale=0.32]{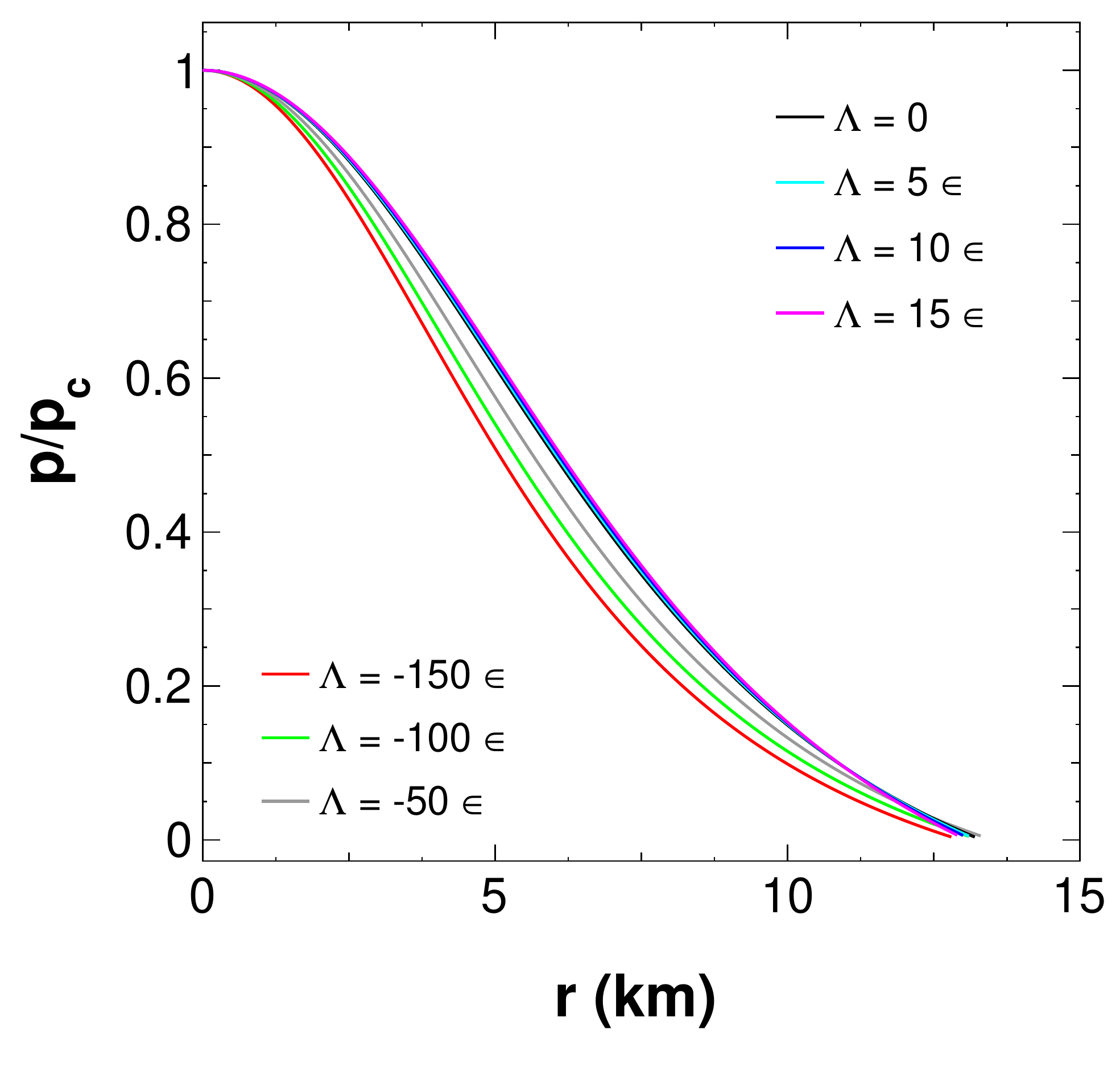}\hspace{0.5cm}
        \includegraphics[scale=0.32]{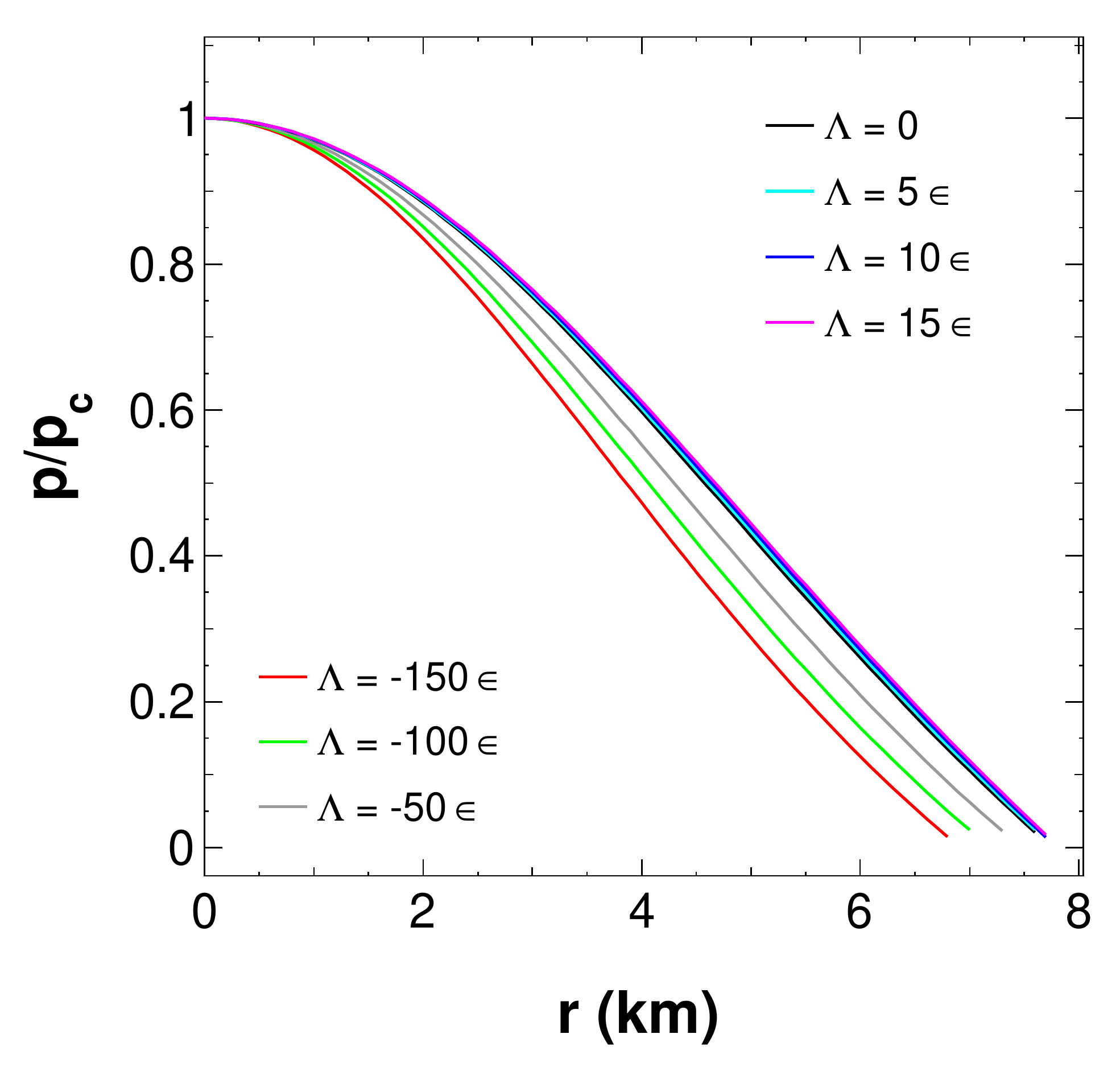}}
        \vspace{-0.2cm}
        \caption{Pressure profiles of strange stars with different 
$\Lambda$ values for the MIT Bag model EoS (first panel) and the linear EoS
(second panel). Here the pressure is scaled in terms of central pressure $p_c$.}
        \label{fig10}
        \end{figure*}
At this point for the clarity about the physical boundary condition that we
have used to calculate the radius of a star as mentioned in section 
\ref{gr}, we have plotted Fig.\ \ref{fig10}, which shows the 
variation of pressure with radial distance $r$ (in km) for different strange 
star configurations as predicted by the MIT Bag model (left panel) and linear 
EoS (right panel). These two pressure profiles indicate that pressure near the 
centre of the star is maximum and it decreases with an increase in the radial 
distance $r$. Hence, at the surface of the star, the pressure becomes zero as 
expected.

\begin{figure*}
	\centerline{
	\includegraphics[scale=0.32]{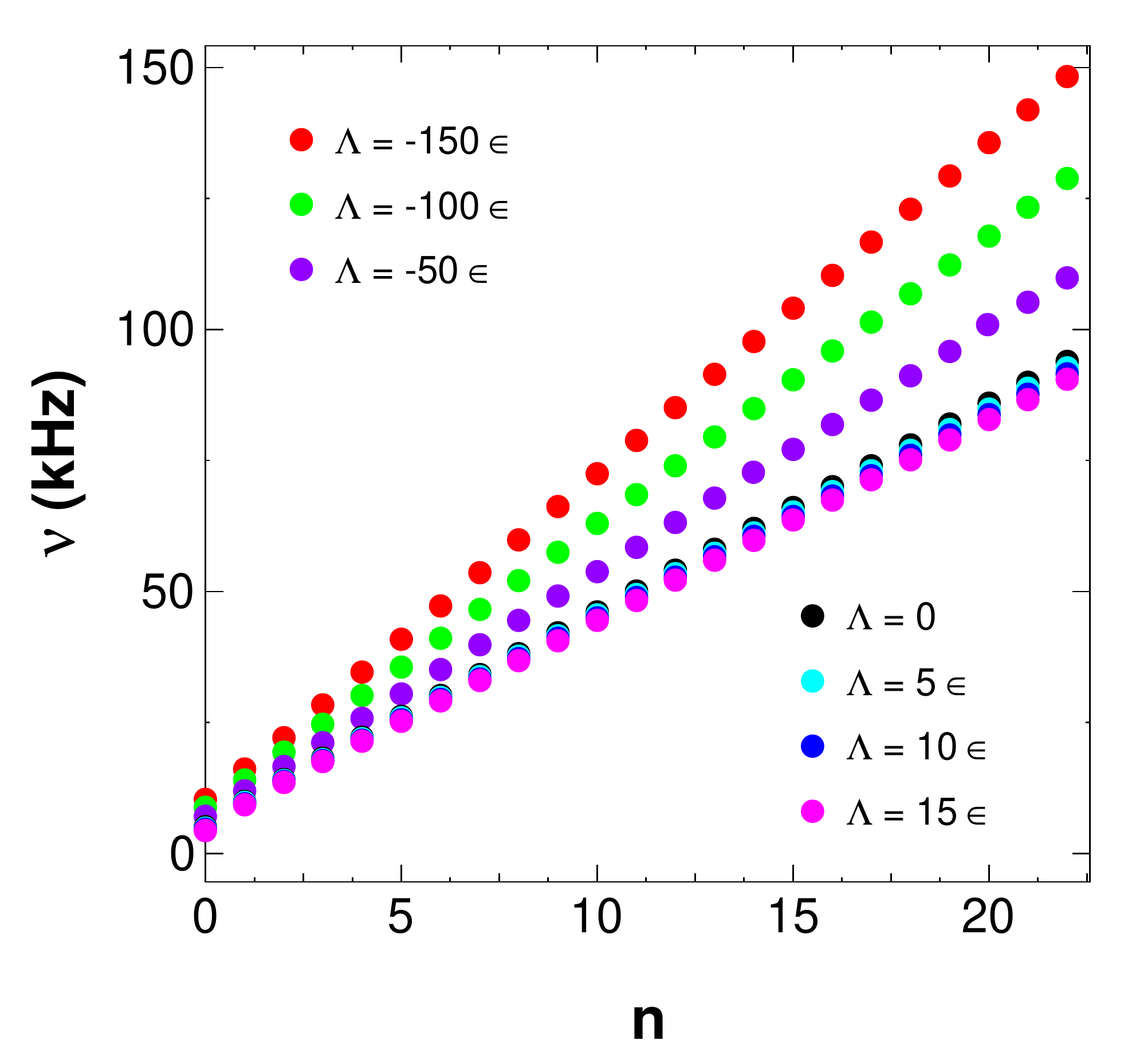}\hspace{0.5cm}
	\includegraphics[scale=0.32]{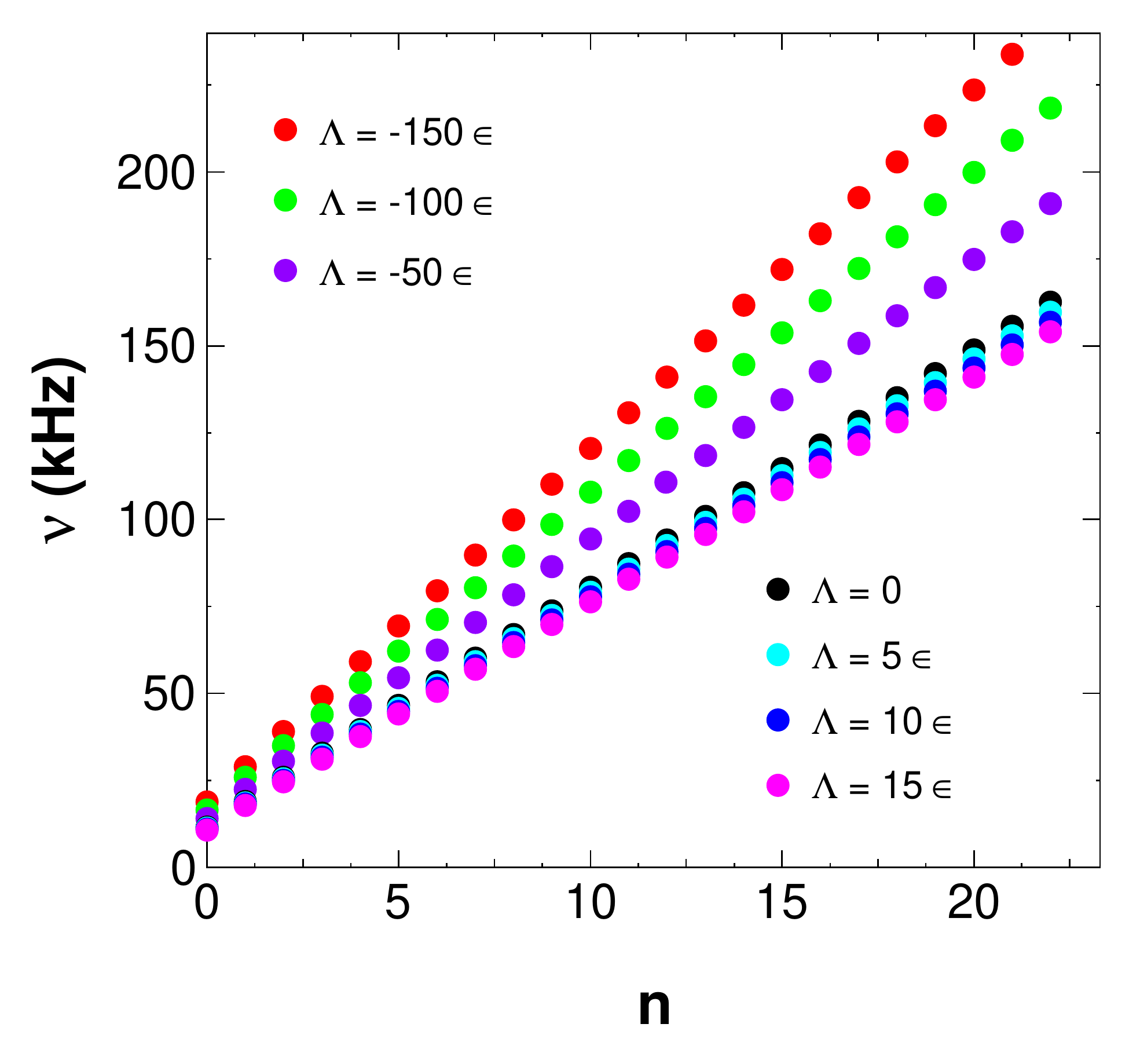}}	
	\vspace{-0.2cm}
	\caption{Variation of radial frequencies $\nu_{n}$ with oscillation 
modes $n$ for different $\Lambda$ values and corresponding to different 
masses and radii which are listed in Tab. \ref{tab:table3}. The first plot is for the MIT Bag 
model EoS and the second plot is for the linear EoS.} 
	\label{fig2} 
	\end{figure*}
As mentioned above, we have calculated the fundamental $f$-mode and 22 lowest 
$p$-modes of radial oscillation of strange stars for the considered EoSs. The 
variation of oscillation frequencies $\nu_{n}$ (in kHz) is plotted against 
the oscillation modes $n$ in Fig.\ \ref{fig2}. Oscillation frequency increases 
linearly with the order of modes. For all values of $\Lambda$, it is increasing 
linearly with increasing modes for both of the EoSs. We have observed the 
maximum oscillation frequency for $\Lambda=-150\,\epsilon$, and the minimum 
frequency is obtained for $\Lambda=15\,\epsilon$ for both of the EoSs. The 
first panel of this figure is for the MIT Bag model EoS and the second panel is 
for the linear EoS. Linear EoS is giving a frequency above $240\,\mbox{kHz}$, 
while for the case of the MIT Bag model the maximum frequency is not exceeding $150\,\mbox{kHz}$. 
\begin{table*}
        \caption{\label{tab:table3} Radial oscillation frequencies $\nu_{n}$
        in kHz for the MIT Bag model EoS for negative, zero and positive
values of $\Lambda$.\vspace{5pt}}
\begin{tabular}{cccccccc}
\hline \hline
Modes (Order n) & $\Lambda=-150\,\epsilon$ & $\Lambda=-100\,\epsilon$ & $\Lambda=-50\,\epsilon$ & $\Lambda=0$ & $
\Lambda=5\,\epsilon$ & $\Lambda=10\,\epsilon$ & $\Lambda=15\,\epsilon$ \\[5pt]
\hline
$f$      (0)  &  10.34 &  8.76  & 7.04    & 5.08   & 4.85   & 4.59   & 4.31    \\
$p_{1}$  (1)  &  16.07 & 14.02  & 11.91   & 9.88   & 9.69   & 9.49   & 9.30    \\
$p_{2}$  (2)  &  22.09 & 19.28  & 16.52   & 14.09  & 13.88  & 13.68  & 13.48   \\
$p_{3}$  (3)  &  28.31 & 24.67  & 21.14   & 18.15  & 17.91  & 17.68  & 17.47   \\
$p_{4}$  (4)  &  34.59 & 30.11  & 25.78   & 22.16  & 21.88  & 21.62  & 21.38   \\
$p_{5}$  (5)  &  40.90 & 35.58  & 30.43   & 26.16  & 25.83  & 25.52  & 25.25   \\
$p_{6}$  (6)  &  47.21 & 41.06  & 35.10   & 30.14  & 29.76  & 29.41  & 29.10   \\
$p_{7}$  (7)  &  53.53 & 46.54  & 39.76   & 34.13  & 33.69  & 33.29  & 32.94   \\
$p_{8}$  (8)  &  59.84 & 52.02  & 44.43   & 38.11  & 37.62  & 37.17  & 36.77   \\
$p_{9}$  (9)  &  66.16 & 57.05  & 49.10   & 42.09  & 41.55  & 41.05  & 40.60   \\
$p_{10}$ (10) &  72.47 & 62.99  & 53.77   & 46.07  & 45.48  & 44.93  & 44.43   \\
$p_{11}$ (11) &  78.79 & 68.47  & 58.45   & 50.06  & 49.41  & 48.81  & 48.27   \\
$p_{12}$ (12) &  85.10 & 73.95  & 63.12   & 54.04  & 53.34  & 52.69  & 52.10   \\
$p_{13}$ (13) &  91.41 & 79.43  & 67.79   & 58.03  & 57.27  & 56.57  & 55.93   \\
$p_{14}$ (14) &  97.71 & 84.91  & 72.47   & 62.01  & 61.20  & 60.44  & 59.76   \\
$p_{15}$ (15) & 104.02 & 90.37  & 77.14   & 66.00  & 65.13  & 64.33  & 63.60   \\
$p_{16}$ (16) & 110.33 & 95.86  & 81.82   & 69.99  & 69.06  & 68.21  & 67.43   \\
$p_{17}$ (17) & 116.65 & 101.36 & 86.49   & 73.98  & 72.99  & 72.09  & 71.26   \\
$p_{18}$ (18) & 122.96 & 106.77 & 91.15   & 77.96  & 76.93  & 75.97  & 75.10   \\
$p_{19}$ (19) & 129.28 & 112.31 & 95.82   & 81.95  & 80.86  & 79.85  & 78.93   \\
$p_{20}$ (20) & 135.60 & 117.81 & 100.50  & 85.94  & 84.79  & 83.74  & 82.77  \\
$p_{21}$ (21) & 141.91 & 123.33 & 105.19  & 89.93  & 88.73  & 87.62  & 86.61  \\
$p_{22}$ (22) & 148.23 & 128.77 & 109.85  & 93.92  & 92.66  & 91.50  & 90.44  \\[1pt]
\hline \hline
\end{tabular}
\end{table*}
In Tables \ref{tab:table3} and \ref{tab:table4}, we summarize the radial 
oscillation frequency modes of strange stars obtained for the MIT Bag model 
EoS and the linear EoS respectively for all considered values of $\Lambda$. 
From these tables, it is inferred that for all these modes, more positive 
$\Lambda$ values are giving smaller oscillation frequencies i.e.\ with 
increasing $\Lambda$ values oscillation frequencies are getting smaller for 
each oscillation mode.
\begin{table*}
        \caption{\label{tab:table4} Radial oscillation frequencies $\nu_{n}$
in kHz for the linear EoS for negative, zero and positive values of $\Lambda
        $.\vspace{5pt}}
\begin{tabular}{cccccccc}
\hline \hline
Modes (Order n) & $\Lambda=-150\,\epsilon$ & $\Lambda=-100\,\epsilon$ & $\Lambda=-50\,\epsilon$ & $\Lambda=0$ & $\Lambda=5\,\epsilon$ & $\Lambda=10\,\epsilon$ & $\Lambda=15\,\epsilon$ \\[5pt]
\hline
$f$      (0)  & 18.67  & 16.44   & 14.03   & 11.44 & 11.18  & 10.91  & 10.64    \\
$p_{1}$  (1)  & 28.90  & 25.78   & 22.42   & 18.86 & 18.49  & 18.13  & 17.76    \\
$p_{2}$  (2)  & 38.98  & 34.90   & 30.52   & 25.89 & 25.41  & 24.94  & 24.47    \\
$p_{3}$  (3)  & 49.06  & 43.97   & 38.52   & 32.79 & 32.21  & 31.63  & 31.05    \\
$p_{4}$  (4)  & 59.17  & 53.04   & 46.50   & 39.64 & 38.94  & 38.25  & 37.56    \\
$p_{5}$  (5)  & 69.32  & 62.12   & 54.47   & 46.46 & 45.65  & 44.84  & 44.04    \\
$p_{6}$  (6)  & 79.51  & 71.22   & 62.44   & 53.27 & 52.35  & 51.42  & 50.50    \\
$p_{7}$  (7)  & 89.72  & 80.35   & 70.42   & 60.08 & 59.03  & 58.00  & 56.96    \\
$p_{8}$  (8)  & 99.95  & 89.49   & 78.41   & 66.88 & 65.72  & 64.57  & 63.41    \\
$p_{9}$  (9)  & 110.21 & 98.64   & 86.41   & 73.69 & 72.41  & 71.14  & 69.87    \\
$p_{10}$ (10) & 120.47 & 107.81  & 94.42   & 80.50 & 79.10  & 77.71  & 76.32    \\
$p_{11}$ (11) & 130.76 & 116.98  & 102.43  & 87.32 & 85.80  & 84.29  & 82.78    \\
$p_{12}$ (12) & 141.05 & 126.17  & 110.45  & 94.13 & 92.50  & 90.86  & 89.24    \\
$p_{13}$ (13) & 151.35 & 135.37  & 118.48  & 100.96 & 99.20 & 97.44  & 95.70    \\
$p_{14}$ (14) & 161.66 & 144.57  & 126.52  & 107.78 & 105.91 & 104.03 & 102.17   \\
$p_{15}$ (15) & 171.97 & 153.78  & 134.56  & 114.61 & 112.61 & 110.62 & 108.63   \\
$p_{16}$ (16) & 182.30 & 162.99  & 142.60  & 121.45 & 119.33 & 117.21 & 115.11   \\
$p_{17}$ (17) & 192.62 & 172.21  & 150.65  & 128.28 & 126.04 & 123.81 & 121.58   \\
$p_{18}$ (18) & 202.95 & 181.43  & 158.70  & 135.12 & 132.76 & 130.40 & 128.06   \\
$p_{19}$ (19) & 213.28 & 190.65  & 166.76  & 141.97 & 139.48 & 137.00 & 134.53   \\
$p_{20}$ (20) & 223.62 & 199.88  & 174.82  & 148.81 & 146.20 & 143.60 & 141.01   \\
$p_{21}$ (21) & 233.96 & 209.11  & 182.88  & 155.66 & 152.93 & 150.21 & 147.50   \\
$p_{22}$ (22) & 244.30 & 218.35  & 190.94  & 162.51 & 159.65 & 156.81 & 153.98   \\[1pt]
\hline \hline
\end{tabular}
\end{table*}
Also, each mode frequency for all $\Lambda$ values is larger for the linear 
EoS than that for the MIT Bag model EoS. These are already visualized in
Fig.\ \ref{fig2}.

\begin{figure*}
	\centerline{
	\includegraphics[scale=0.32]{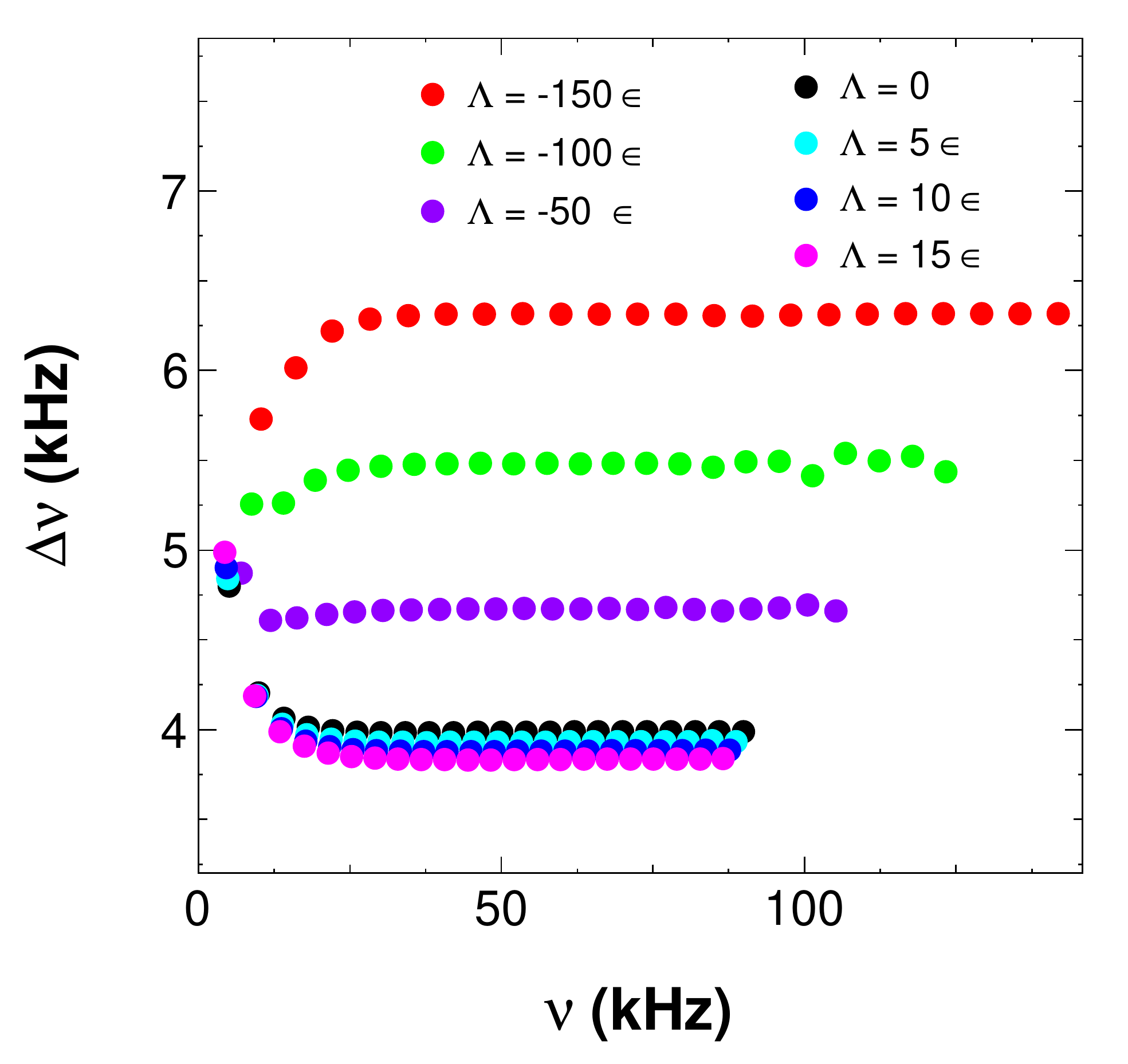}\hspace{0.5cm}
	\includegraphics[scale=0.32]{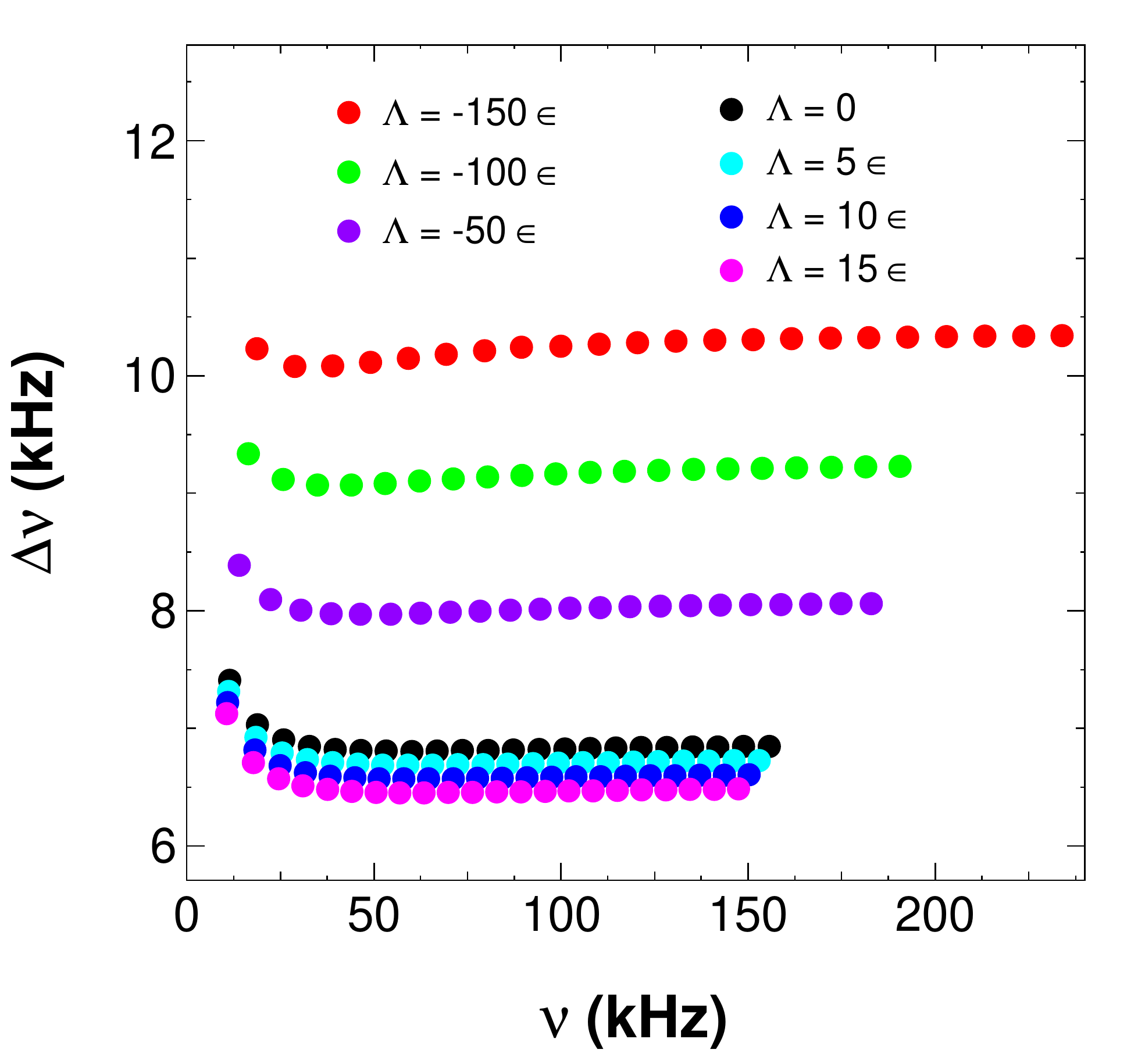}}	
	\vspace{-0.2cm}
	\caption{Variation of difference between consecutive modes of radial 
frequencies ($\Delta\nu_{n}$) with radial frequencies $\nu_n$ for different 
$\Lambda$ values. First panel and second panel are for the MIT Bag model EoS 
and linear EoS respectively.} 
	\label{fig3} 
	\end{figure*}
In Fig.\ \ref{fig3}, the variation of separation between two consecutive 
modes of radial frequencies ($\Delta\nu$) with respect to the mode frequencies 
$\nu$ is shown. The first panel is for the MIT Bag model and the second panel 
is for the linear EoS. For the MIT Bag model with positive $\Lambda$ values, a 
gradually decreasing pattern is observed. The difference between the 
oscillation frequencies of $f$-mode and $p_{1}$-mode is larger than that of 
the other values. However, in the case of more negative values of $\Lambda$, 
the opposite behaviour is observed. Unlike other $\Lambda$ values, for 
$\Lambda=-150\,\epsilon$ and $-100\,\epsilon$ the difference between 
oscillation frequencies of $f$-mode and $p_{1}$-mode is smaller than the 
rest of the variations. In the case of linear EoS, nearly a smooth variation 
for all the considered $\Lambda$ values is observed. For 
$\Lambda=-150\,\epsilon$ both EoSs give maximum value of $\Delta\nu$ and
for $\Lambda= 15\,\epsilon$ they give a minimum value of $\Delta\nu$
in case of modes other than $f$-mode and $p_{1}$-mode with MIT Bag model EoS.
That is under this condition $\Delta\nu$ decreases with increasing values of
$\Lambda$. This figure also clarifies that for $\Lambda = -150\,\epsilon$, the 
radial frequencies are maximum whereas the frequencies are minimum for $\Lambda = 
15\,\epsilon$ for both MIT Bag model EoS and linear EoS  as also
seen from Tables \ref{tab:table3} and \ref{tab:table4} respectively.

\begin{figure*}
	\centerline{
	\includegraphics[scale=0.32]{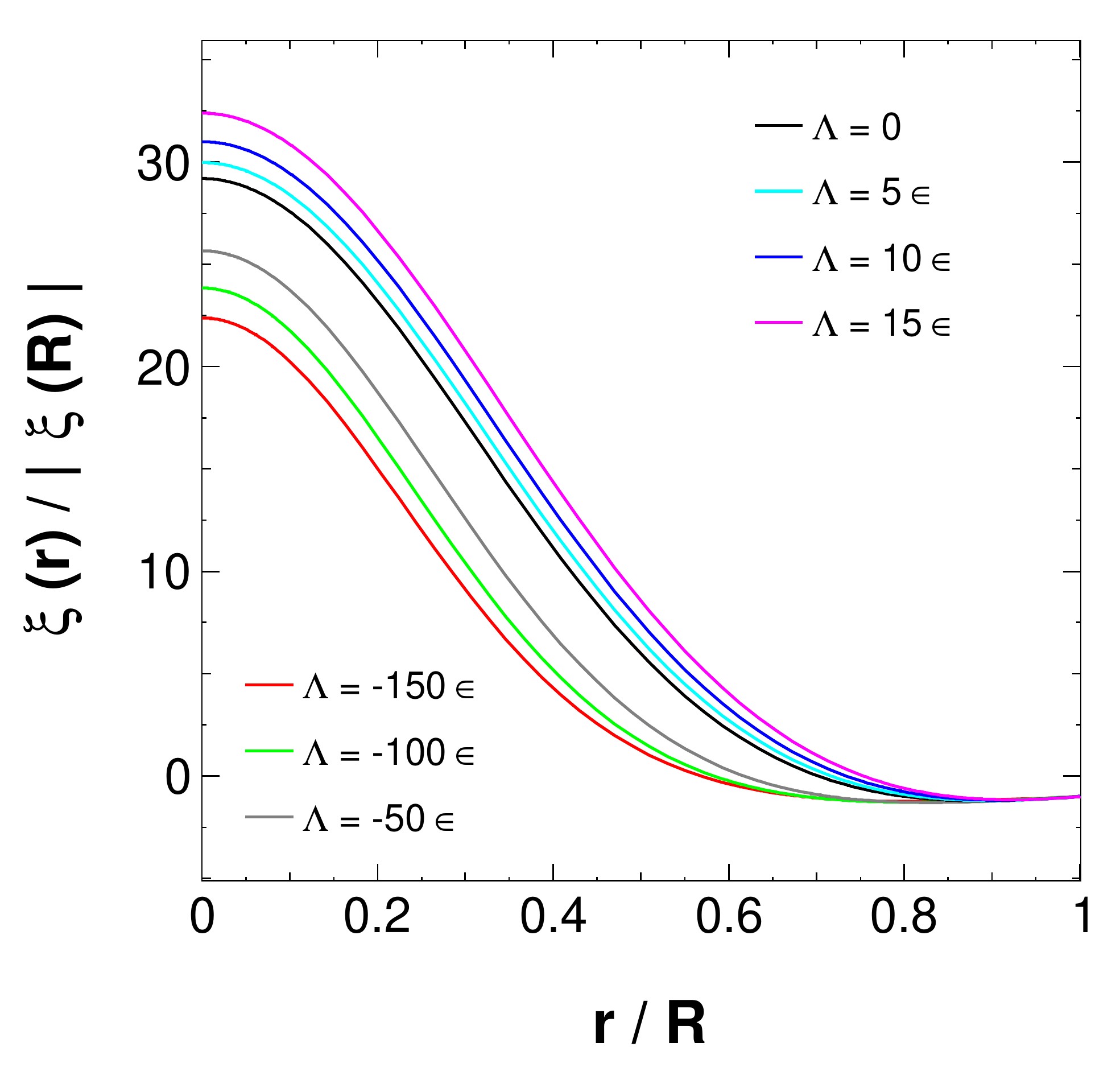}\hspace{0.5cm}
	\includegraphics[scale=0.32]{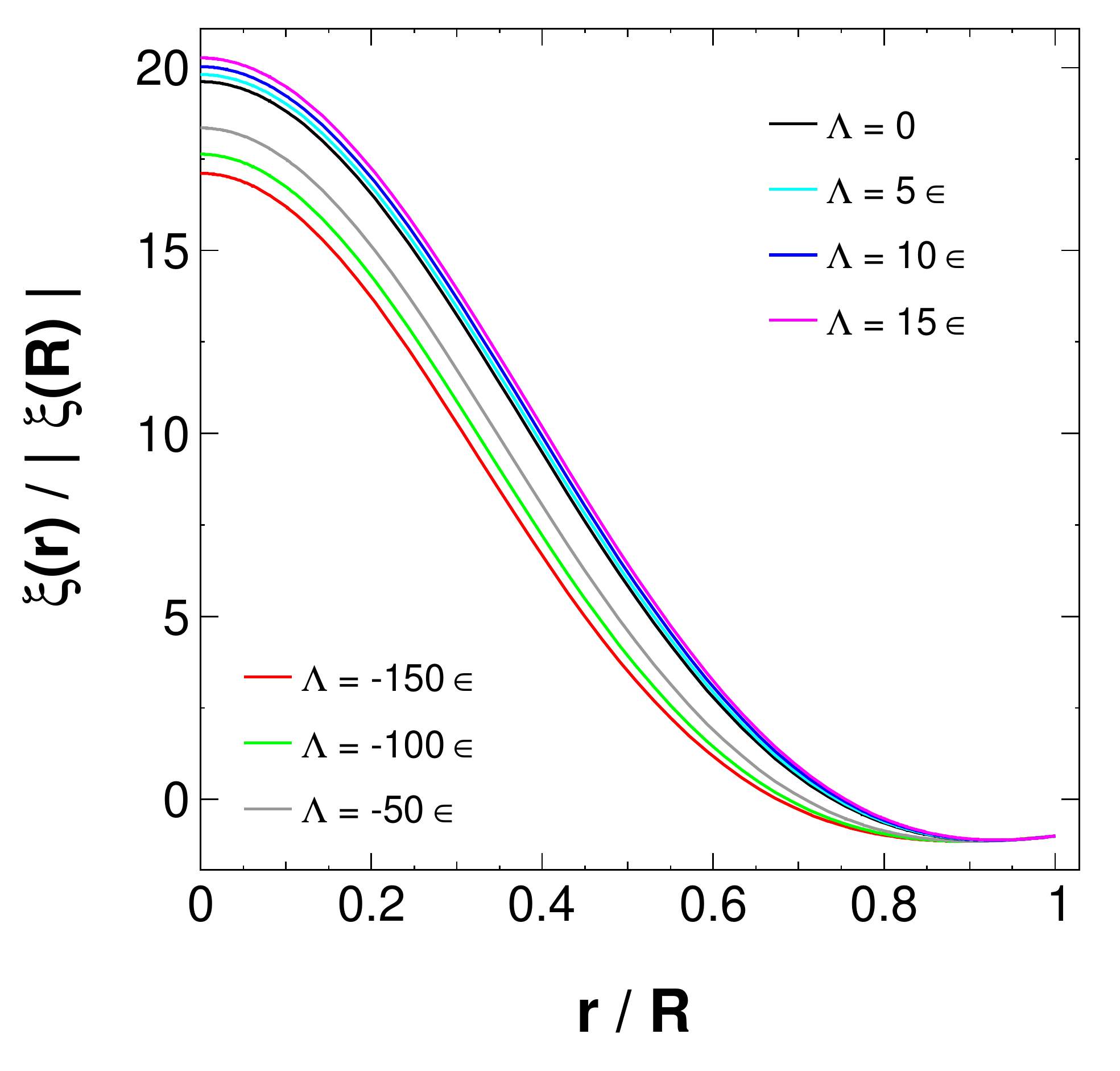}}	
	\vspace{-0.3cm}
	\caption{Variation of $f$-mode of radial perturbations $\xi(r)$ with 
radial distance $r$ (in km) obtained by using the MIT Bag model EoS (first 
panel) and the linear EoS (second panel) for different $\Lambda$ values. } 
	\label{fig4} 
	\end{figure*}
\begin{figure*}
	\centerline{
	\includegraphics[scale=0.32]{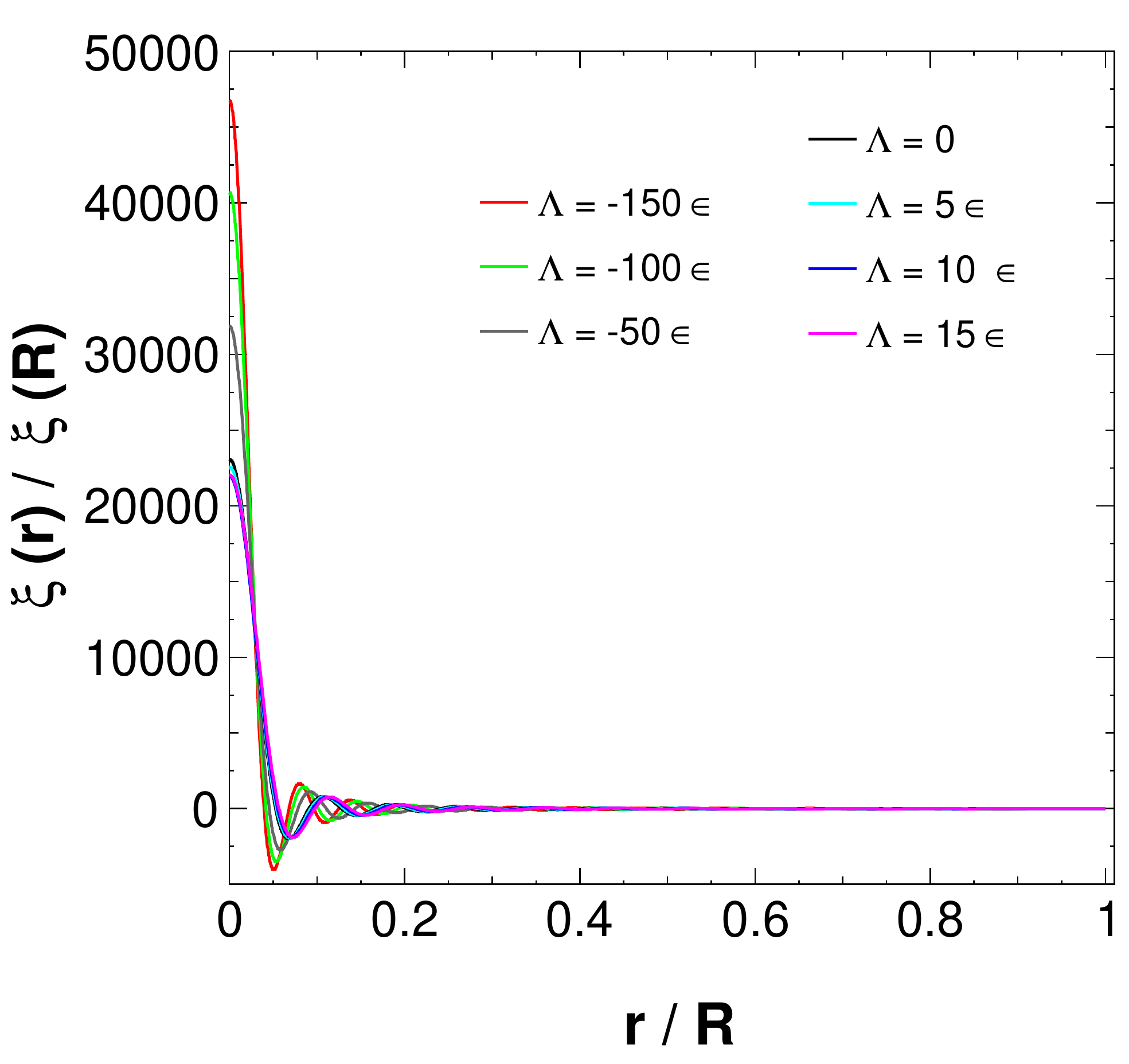}\hspace{0.5cm}
	\includegraphics[scale=0.32]{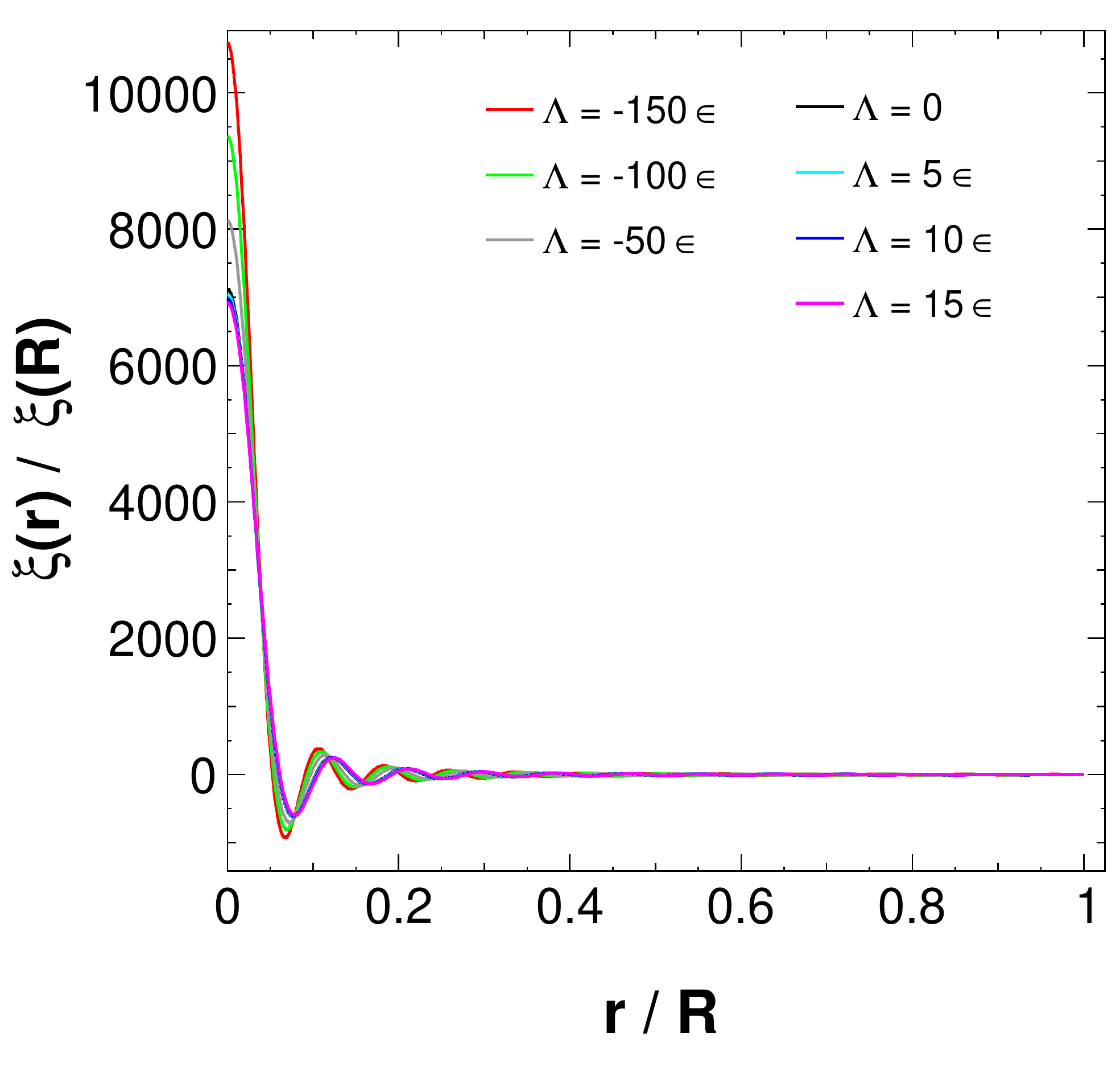}}	
	\vspace{-0.2cm}
	\caption{Variation of radial perturbations $\xi(r)$ with radial 
distance $r$ (in km) obtained by using the MIT Bag model EoS (first panel) and 
the linear EoS (second panel) for different $\Lambda$ values for the case of 
pressure $p_{22}$-mode.} 
	\label{fig5}
	\end{figure*}

From the Sturm-Liouville dynamic pulsation equations in presence of a 
cosmological constant (equations \eqref{eq9}-\eqref{eq10}), we have studied 
the variation of radial and pressure perturbations with the radial distance 
$r$. In Fig.\ \ref{fig4} and \ref{fig5} the variation of radial perturbation 
$\xi(r)$ with radial distance $r$ is shown for $f$-mode and $p_{22}$-mode 
respectively. In the first panel of Fig.\ \ref{fig4}, the variation of radial 
perturbation $\xi(r)$ with $r$ (in km) is shown for the MIT Bag model and in 
the second panel, the variation is shown for the linear EoS. From these 
two plots, it is clear that the radial perturbation is larger only near the 
centre of the star. It is diminishing near the surface of the star for both 
these EoSs. As the EoSs with different $\Lambda$ values are mimicking strange 
stars of different sizes, the perturbation along the radial distances are 
found different.  

Similarly in Fig.\ \ref{fig5}, the variation is shown for higher-order 
$p_{22}$-mode. For the MIT Bag model, it is shown in the first panel of the 
figure. As like the $f$-mode, the perturbation is larger near the centre of 
the star. Towards the surface, a distinct damping nature of radial perturbation 
is noticed. The different $\Lambda$ values are also showing distinct variations 
in their respective perturbations. As shown in the second panel of Fig.\ 
\ref{fig5}, for the case of linear EoS, the variation of $p_{22}$ mode is 
behaving similarly to that of the MIT Bag model EoS. The perturbation is 
decreasing along the surface of the star. Intermediate pressure perturbations 
curves can be drawn which will lie in between $f$ and $p_{22}$ modes.

\begin{figure*}
	\centerline{
	\includegraphics[scale=0.32]{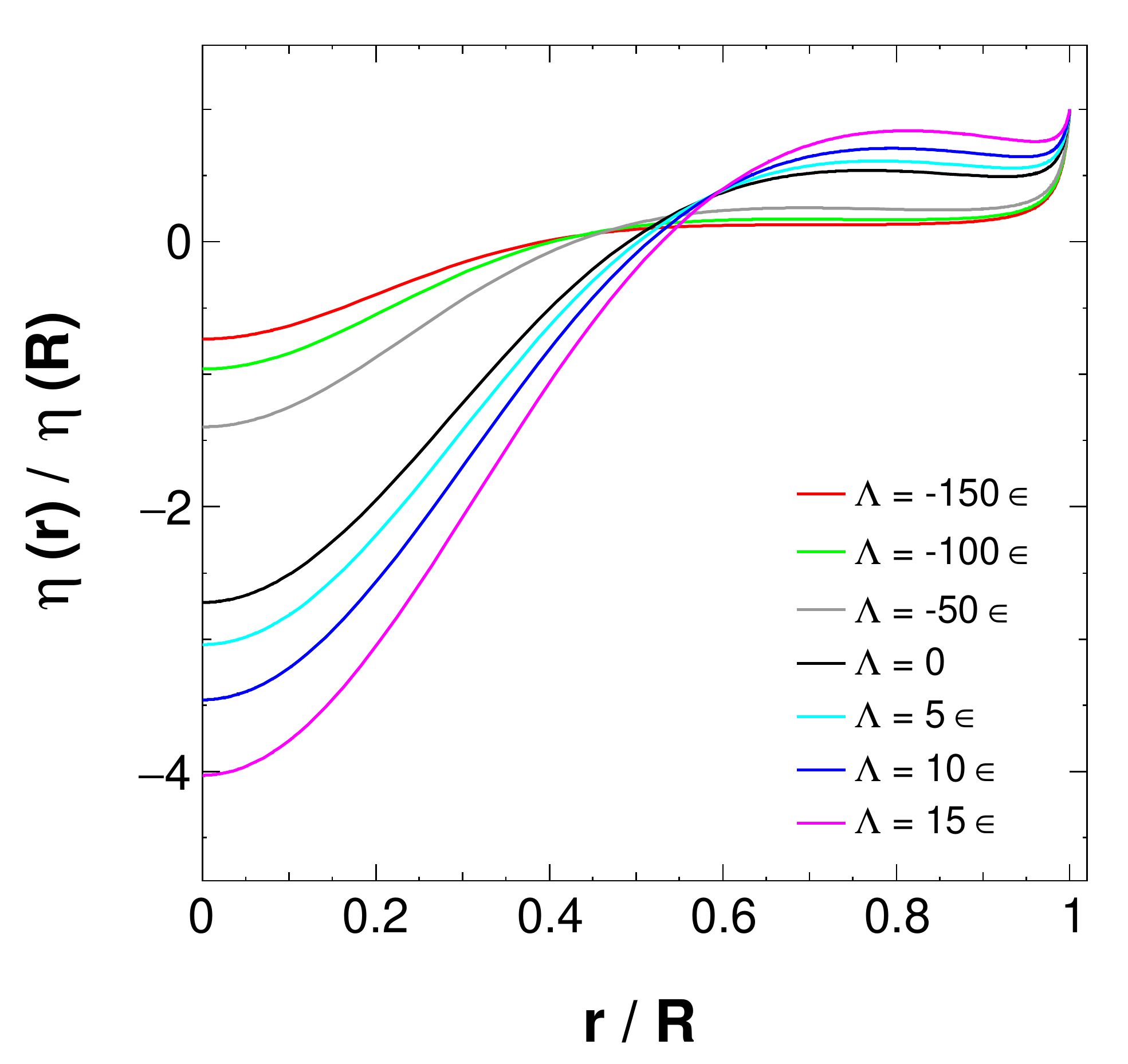}\hspace{0.5cm}
	\includegraphics[scale=0.32]{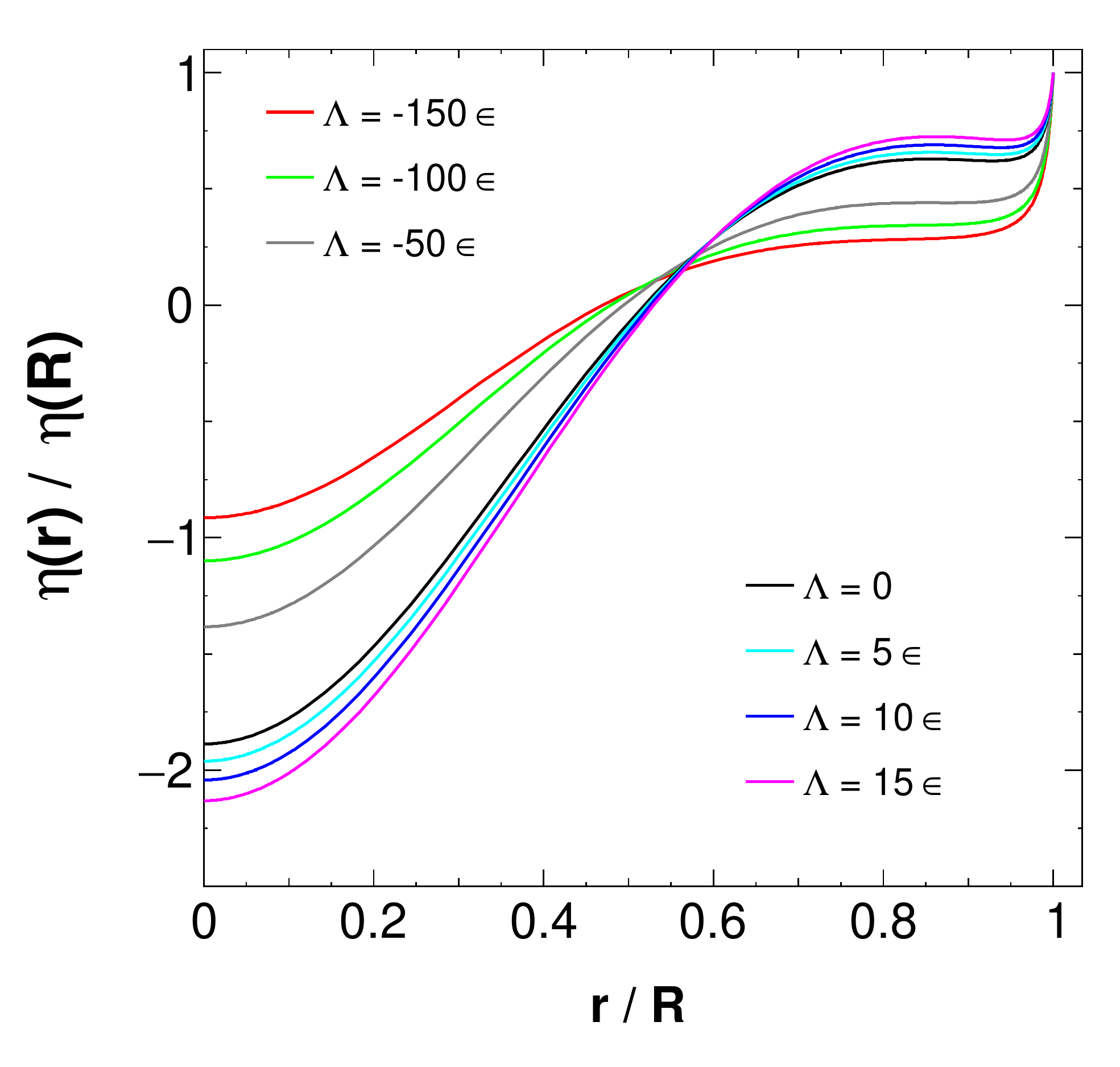}}	
	\vspace{-0.2cm}
	\caption{Variation showing pressure perturbations $\eta(r)$ with 
radial distance $r$ (in km) found by using MIT the Bag model EoS (first panel) 
and the linear EoS (second panel) for different $\Lambda$ values for 
fundamental $f$-mode.} 
	\label{fig6} 
	\end{figure*}
\begin{figure*}
	\centerline{
	\includegraphics[scale=0.32]{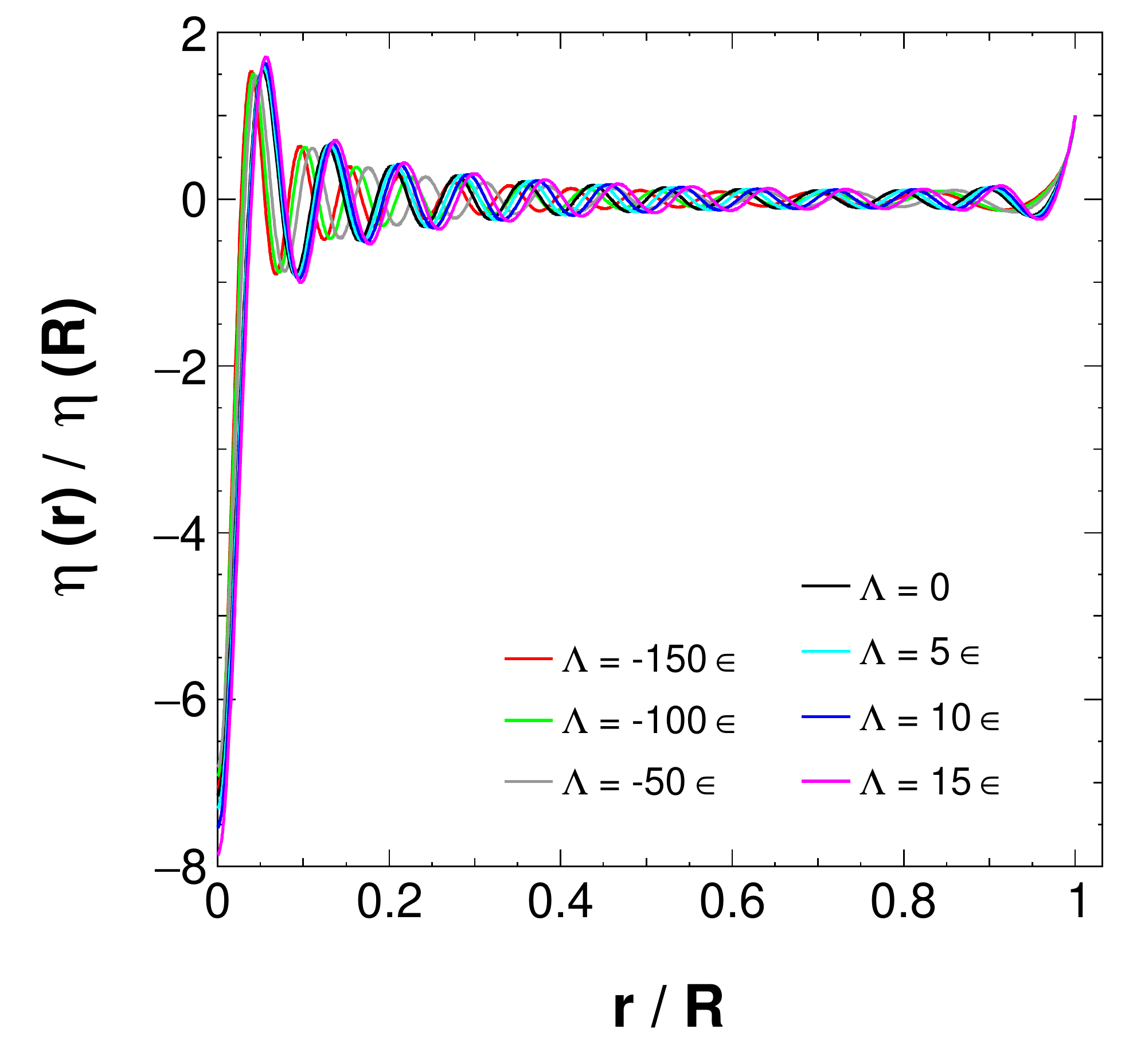}\hspace{0.5cm}
	\includegraphics[scale=0.32]{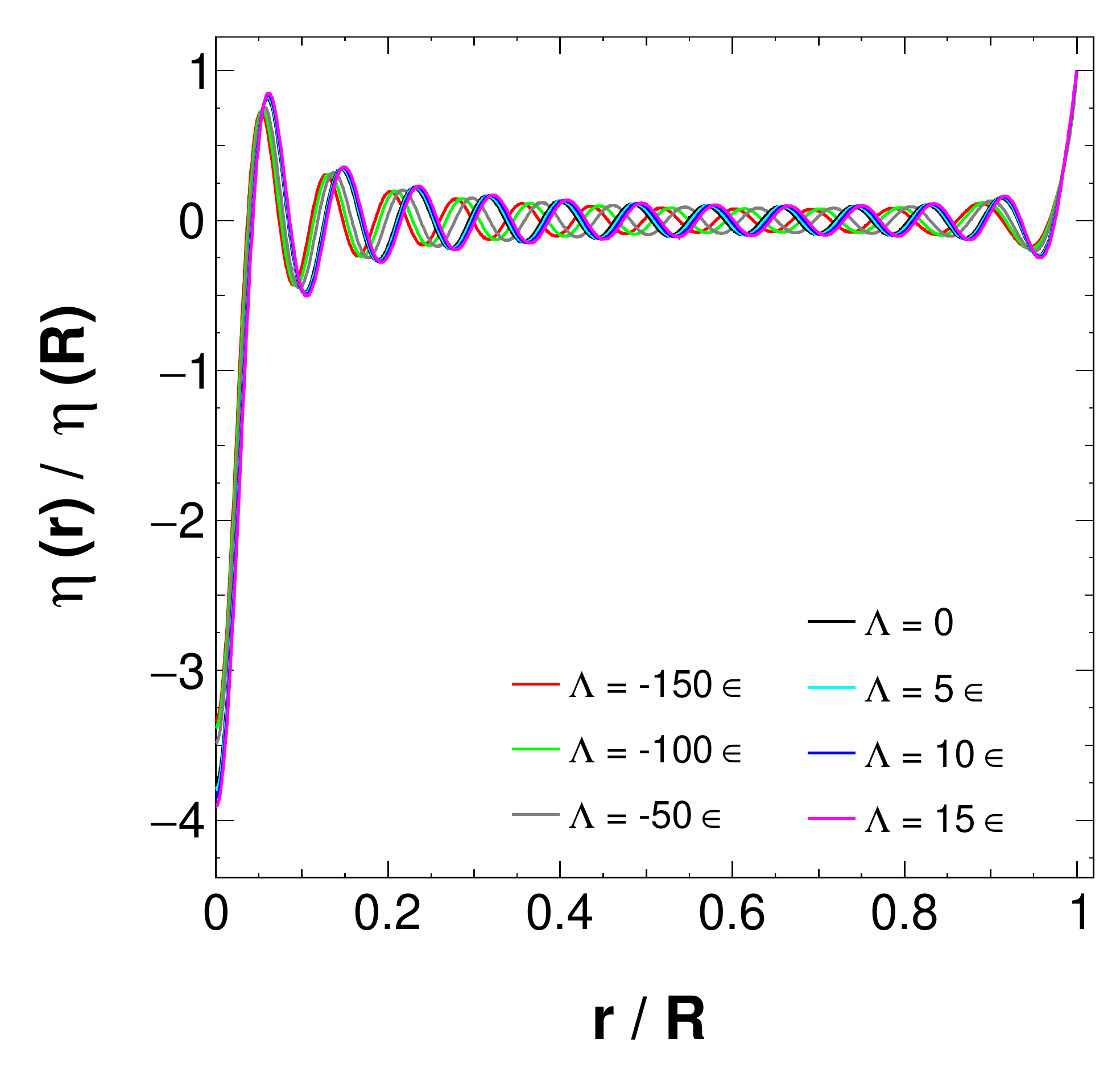}}	
	\vspace{-0.2cm}
	\caption{Variation showing pressure perturbations $\eta(r)$ with 
radial distance $r$ (in km) found by using the MIT Bag model EoS (first panel) 
and the linear EoS (second panel) for different $\Lambda$ values for the 
pressure $p_{22}$-mode.} 
	\label{fig7} 
	\end{figure*}

The variation of pressure perturbation is different from the variation of 
radial perturbation. As clear from Fig.\ \ref{fig6} and \ref{fig7}, the 
variation is larger near the centre and surface of each star. Whereas, for 
the case of radial perturbation, it is larger near the centre of the star 
only. In the first panel of Fig.\ \ref{fig6}, the change in pressure 
perturbation with distance from the centre of the star is showing for the MIT 
Bag model EoS for fundamental $f$-mode due to different $\Lambda$ values. The 
different sizes of each star for each $\Lambda$ values are also clear from 
this plot. Similar behaviour is observed for linear EoS with fundamental 
$f$-mode. This can be visualised from the second panel of this Fig.\ \ref{fig6}.These variations shown in this figure correspond to different values 
of mass and radius corresponding to considered $\Lambda$ values. As an example, 
for the MIT Bag model EoS i.e. in the left panel of Fig.\ \ref{fig6} and for $\Lambda=15\,\epsilon$, the corresponding mass and radius are $\approx 3.3\,\mbox{M}_{\odot}$ and $\approx 14.3\,\mbox{km}$ respectively. The other values 
of mass and radius corresponding to these variations can be found in Table \ref{tab:table1}.

This pressure variation for higher-order oscillation mode i.e.\ for 
$p_{22}$-mode is shown in Fig.\ \ref{fig7}. The first panel corresponds 
to the MIT Bag 
model and the second panel corresponds to the linear EoS. Similar to the 
$f$-mode of oscillation this variation is also larger near the centre and near 
the surface of the star. For all other modes in between $f$-mode and 
$p_{22}$-mode, pressure perturbation will stay in between.

\begin{figure*}
	\centerline{
	\includegraphics[scale=0.32]{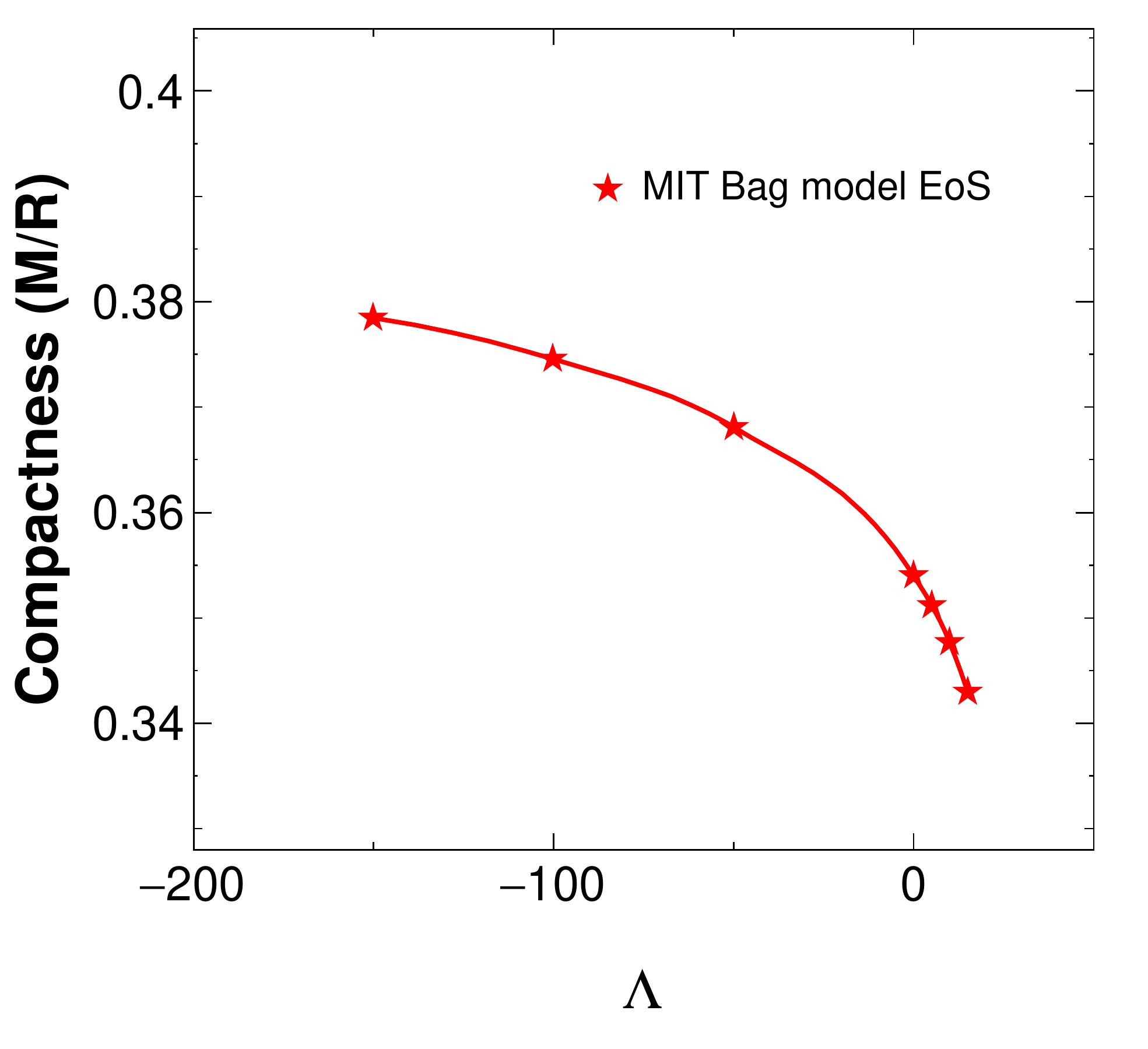}\hspace{0.5cm}
	\includegraphics[scale=0.32]{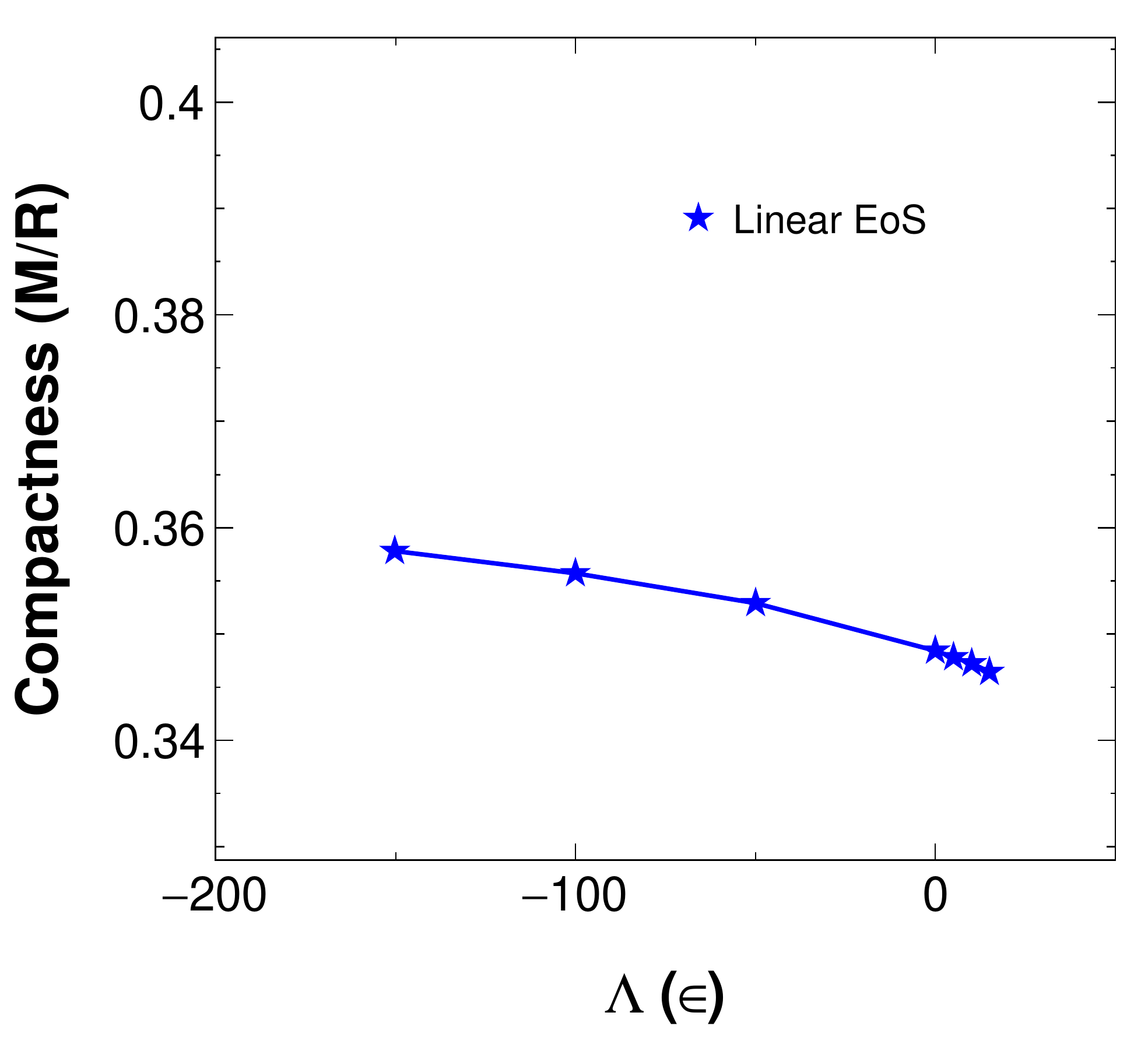}}	
	\vspace{-0.2cm}
	\caption{The variation of compactness of strange stars with different $
	\Lambda$ values for the MIT Bag model EoS (first panel) and the linear EoS 
(second panel).} 
	\label{fig9} 
	\end{figure*}
The value of cosmological constant $\Lambda$ varying with different strange 
star configurations can be visualized in Fig. \ref{fig9}. The left panel of 
this figure is for the MIT Bag model and the right panel is for the linear EoS.
For both of the cases, with increasing $\Lambda$ value compactness is found to 
decrease. The fall is somewhat rapid for the MIT Bag model EoS than that of the
linear EoS. The different values of these plots can be found in Table 
\ref{tab:table3} and in Table \ref{tab:table4}.  

\begin{figure*}
	\centerline{
	\includegraphics[scale=0.32]{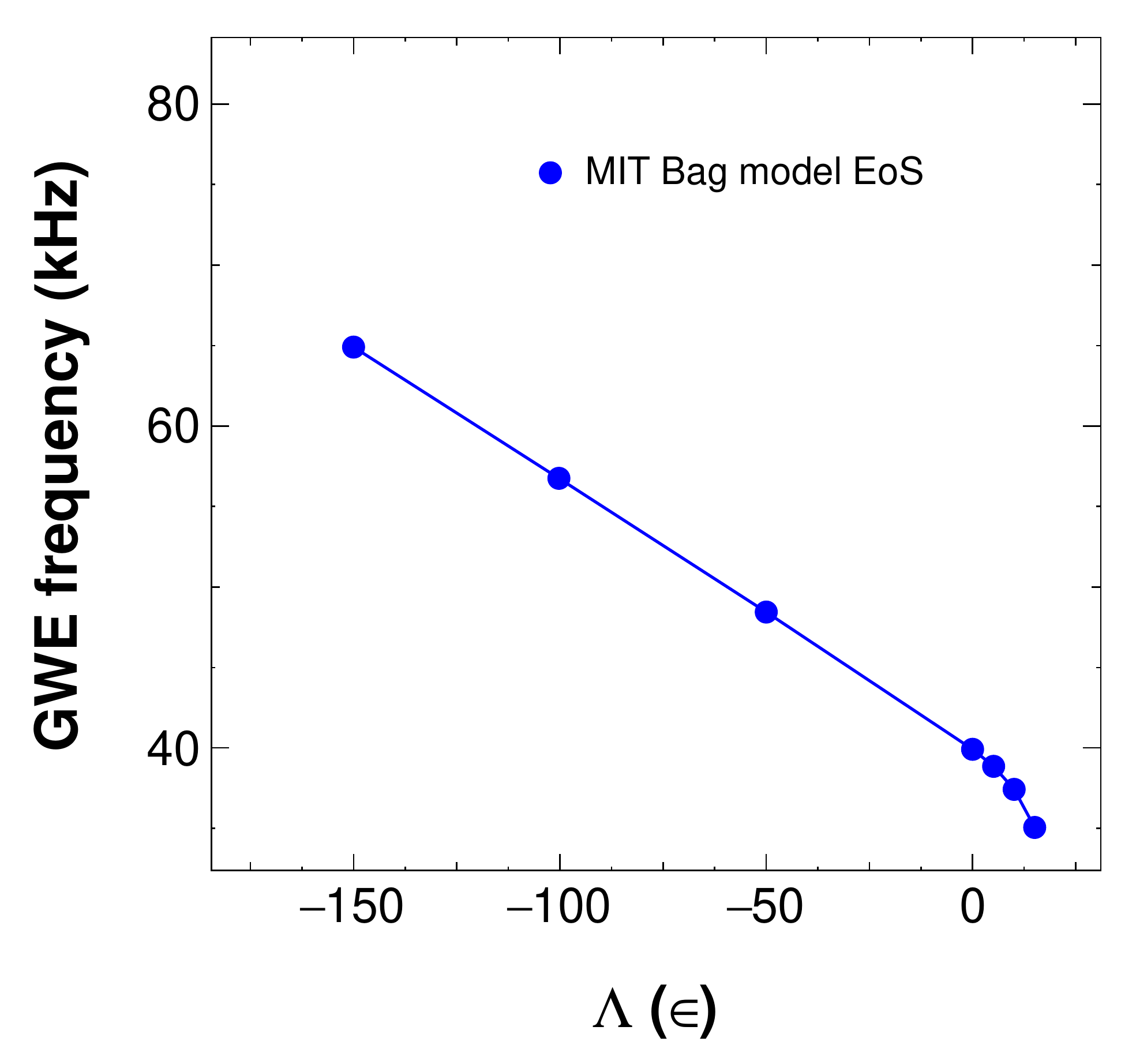}\hspace{0.5cm}
	\includegraphics[scale=0.32]{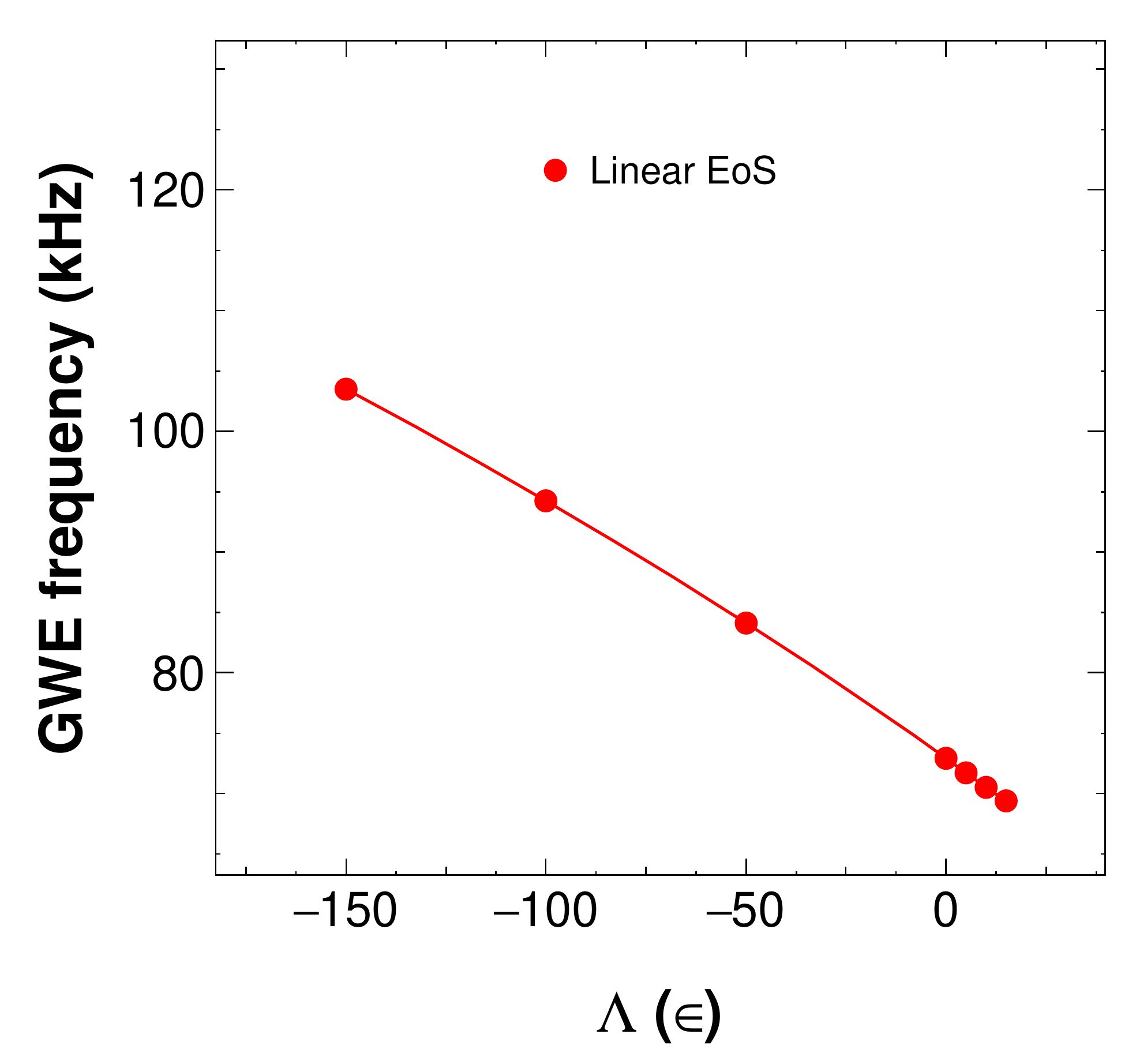}}	
	\vspace{-0.2cm}
	\caption{The variation of GWE frequencies with different $\Lambda$ 
values for the MIT Bag model EoS (first panel) and the linear EoS 
(second panel).} 
	\label{fig8} 
	\end{figure*}
After the numerical analysis of the perturbation equations, another 
important part of this work is to find the effect of cosmological constant 
$\Lambda$ on GWE frequencies. The GWE frequencies are found to diminish with an
increase in $\Lambda$ values, for both of the EoSs. For the MIT Bag model with 
negative $\Lambda$ values the variation is straightaway, however, a rapid drop 
is observed for positive $\Lambda$ values. The linear EoS is showing almost a 
straight variation for both positive and negative values of $\Lambda$. These 
variations are shown in Fig.\ \ref{fig8}, where the first panel is 
for the MIT Bag model and the second panel is for the linear EoS. 

\begin{table*}
\caption{\label{tab:table1} Mass, radius, compactness of  strange stars, 
characteristic echo time and GWEs from  strange stars for the MIT Bag model 
EoS with the bag constant $B=(190\,\mbox{MeV)}^4$.}
\begin{ruledtabular}
\begin{tabular}{cc|ccccc} 
&\hspace{-3cm}Value of $\Lambda$ & Radius R & Mass M & Compactness & Echo & GWE\\ [-2pt] 
(in terms of $\epsilon$) & (in $\,\mbox{cm}^{-2}$) & (in km) & (in $\mbox{M}_{\odot}$) & (M/R) & time (ms) & frequency (kHz)\\ \hline 
-150$\,\epsilon$ &  -2.66$\times10^{-12}$& 13.003 & 3.327 & 0.3784 & 0.048 & 64.90 \\
-100$\,\epsilon$ & -1.77$\times10^{-12}$ & 13.090 & 3.315 & 0.3745 & 0.055 & 56.74 \\
-50$\,\epsilon$  & -8.88$\times10^{-13}$ & 13.259 & 3.300 & 0.3681 & 0.065 & 48.44 \\
0    & 0 & 13.766 & 3.295 & 0.3540 & 0.078 & 39.91 \\
5$\,\epsilon$    & 8.88$\times10^{-14}$ & 13.893 & 3.300 & 0.3512 & 0.080 & 38.86 \\
10$\,\epsilon$   & 1.77$\times10^{-13}$  & 14.067 & 3.308 & 0.3477 & 0.083 & 37.42 \\ 
15$\,\epsilon$   & 2.66$\times10^{-13}$  & 14.340 & 3.326 & 0.3430 & 0.089 & 35.08 \\ 
\end{tabular}
\end{ruledtabular}
\end{table*}
\begin{table*}
\caption{\label{tab:table2} Mass, radius, compactness of strange stars, 
characteristic echo time and GWEs from  strange stars for the linear EoS 
with the linear constant $b=0.910$.}
\begin{ruledtabular}
\begin{tabular}{cc|ccccc} 
&\hspace{-3cm}Value of $\Lambda$ & Radius R & Mass M & Compactness & Echo & GWE\\ [-2pt] 
(in terms of $\epsilon$) & (in $\,\mbox{cm}^{-2}$)& (in km) & (in $\mbox{M}_{\odot}$) & (M/R) & time (ms) & frequency (kHz)\\ \hline  
-150$\,\epsilon$ & -2.66$\times10^{-12}$ & 7.200 & 1.742 & 0.3577 & 0.030 & 103.42 \\
-100$\,\epsilon$ & -1.77$\times10^{-12}$ & 7.271 & 1.749 & 0.3557 & 0.033 &  94.24 \\
 -50$\,\epsilon$ & -8.88$\times10^{-13}$ & 7.372 & 1.759 & 0.3529 & 0.037 &  84.12 \\ 
0   & 0 & 7.535 & 1.775 & 0.3484 & 0.043 &  72.91 \\
 5$\,\epsilon$ & 8.88$\times10^{-14}$ & 7.558 & 1.778 & 0.3478 & 0.044 &  71.71 \\
10$\,\epsilon$ & 1.77$\times10^{-13}$ & 7.584 & 1.780 & 0.3472 & 0.045 &  70.51 \\ 
15$\,\epsilon$ & 2.66$\times10^{-13}$ & 7.610 & 1.783 & 0.3464 & 0.045 &  69.40 \\ 
\end{tabular}
\end{ruledtabular}
\end{table*}
The properties of stellar structure such as mass, radius, and compactness along 
with the results for the GWE frequencies and characteristic echo times that are
obtained by using the MIT Bag model EoS are compiled in Table \ref{tab:table1}. 
From this table, the change in these properties of strange stars for the MIT 
Bag model with different $\Lambda$ values can be seen clearly. With increasing
$\Lambda$ values the size of stars is increasing. The compactness of the stars 
is decreasing with more positive values. Thus the upper limit on cosmological 
constant $\Lambda$ value is depicting a star with compactness such that it 
is just able to echo the falling GWs. For our considered maximum value of 
$\Lambda$ (15$\,\epsilon$), the compactness is found to be 0.3430, which is 
greater than 1/3. The characteristic echo time obtained for this EoS increases 
with the increasing $\Lambda$ value, and hence the reverse effect is noticed 
for the GWE frequencies. For this EoS, all echo frequencies obtained are 
in the range of tens of kilohertz.

The sizes of stars obtained by using the linear EoSs are smaller than that 
obtained for the MIT Bag model EoSs. Eventually, the masses are also small in 
comparison to that for the MIT Bag model and hence fulfilling the criteria 
for echoing GWs. These results are shown in Table \ref{tab:table2}. For all 
the chosen $\Lambda$ values, the stars are found to be with enough 
compactness to emit GWE frequencies. The characteristic echo times obtained 
are smaller than that of the MIT Bag model EoS and hence in turn the larger 
echo frequencies are obtained for this EoS. 

\begin{table*}
\caption{\label{tab:table5} Properties of strange stars for the MIT Bag model and linear EoS with different values of $\Lambda$.}
\begin{ruledtabular}
\begin{tabular}{cc|cccc|cccc} 
&\hspace{-2cm}Value of $\Lambda$ &&& MIT Bag model &&&& Linear EoS&\\  
(in $\,\mbox{cm}^{-2}$) & (in terms of $\epsilon$) & Radius & Mass & Compactness & Echo freq- & Radius & Mass & Compactness & Echo freq-\\ [-3pt] 
& & (in km) &  (in $\mbox{M}_{\odot}$) & (M/R) & uency (kHz) 
& (in km) &  (in $\mbox{M}_{\odot}$) & (M/R) & uency (kHz)\\ \hline 
$1\times10^{-18}$ & 5.6$\times10^{-5}\,\epsilon$ & 13.766 & 3.295 & 0.354 & 39.91 & 7.535 & 1.775 & 0.348 & 72.91 \\
$1\times10^{-16}$ & 5.6$\times10^{-3}\,\epsilon$ & 13.766 & 3.295 & 0.354 & 39.91 & 7.535 & 1.775 & 0.348 & 72.91 \\
$1\times10^{-15}$ & 5.6$\times10^{-2}\,\epsilon$ & 13.767 & 3.296 & 0.354 & 39.90 & 7.535 & 1.775 & 0.348 & 72.89 \\ 
$1\times10^{-14}$ & 5.6$\times10^{-1}\,\epsilon$ & 13.778 & 3.296 & 0.354 & 39.80 & 7.538 & 1.775 & 0.348 & 72.76 \\ 
$5\times10^{-14}$ & 2.8$\,\epsilon$ & 13.834 & 3.298 & 0.353 & 39.34 & 7.548 & 1.776 & 0.348 & 72.23 \\ 
$1\times10^{-13}$ & 5.6$\,\epsilon$ & 13.912 & 3.301 & 0.351 & 38.66 & 7.558 & 1.778 & 0.348 & 71.71 \\ 
$2\times10^{-13}$ & 11.2$\,\epsilon$ & 14.122 & 3.312 & 0.347 & 36.99 & 7.590 & 1.781 & 0.347 & 70.20 \\ 
$3\times10^{-13}$ & 16.8$\,\epsilon$ & 14.492 & 3.338 & 0.341 & 35.26 & 7.621 & 1.784 & 0.346 & 68.28 \\ 
\end{tabular}
\end{ruledtabular}
\end{table*}
	
\section{Conclusions}\label{conclusion} 
In this study, we look into the role of the cosmological constant on two 
interesting aspects of strange stars. In one part we have tried to understand 
its impact over the radial oscillations of strange stars and in another, we 
have investigated its effect on GWE frequencies emitted by a strange star, like 
the star formed in the binary merging event GW170817. This study is made in the 
general relativistic framework. To study its role over radial oscillations we 
solved the TOV equations for two EoSs, viz., the MIT bag model EoS and the 
linear EoS. The solution of TOV equations leads us to know 
the structure of strange stars in presence of cosmological constant $\Lambda$, 
which is found to be different from the structure obtained by using the 
vanishing $\Lambda$ \cite{jb}. Again as discussed in Sec.\ \ref{numerical}, 
different $\Lambda$ values give us different strange star configurations. 
Moreover, the pulsation equations developed by Chandrasekhar are modified 
for spacetime with a cosmological constant by introducing two dimensionless 
parameters $\xi$ and $\eta$. The solution of these pulsation equations gave 
us eigenfrequencies of oscillations and hence we have calculated the $f$-mode 
and first 22 $p$-modes of oscillations. We found that the large value of 
$\Lambda$ decreases the radial oscillation frequencies of stars for both of the 
EoSs. Further, to see the role of $\Lambda$ on GWEs first we calculated the 
characteristic echo time of GWs falling on strange stars. 

The study clearly shows that at the present cosmological constant value around 
$\Lambda=(4.24\, \pm\, 0.11)\times 10^{-66}\, \mbox{eV}^{2}$ ($\equiv\; \sim 
10^{-56}$ cm$^{-2}$), the maximum mass and radius, radial oscillations, and 
echo frequencies of strange stars are identical with the GR values and 
can't be distinguished both theoretically and experimentally. Hence, the 
observations of such parameters of a strange star can't be used to 
differentiate between GR and the $\Lambda$CDM model for the present value of 
$\Lambda$. However, bounds on $\Lambda$ from the stellar stability and 
interior structure \cite{nayak,Bordbar,Liu,arbanil} are comparatively weak, 
which gives us a wide range of possible $\Lambda$ values. We have shown that 
within this boundary, the cosmological constant $\Lambda$ with a large 
magnitude can effectively put significant impacts on the mass, radius, radial 
oscillations, and echo frequencies of a strange star. Our results showed that 
for strange stars, the effective upper limit of $\Lambda$ is 
$\sim3\times10^{-13}\, \mbox{cm}^{-2}$ and the lower limit is 
 $\sim10^{-15}\, \mbox{cm}^{-2}$. The maximum mass, radius, radial oscillation 
frequencies, and echo frequencies of strange star configurations have changed 
significantly within these boundary values. These impacts again depend on the 
EoS of the star. These results are presented in Table \ref{tab:table5}. 
According to the results of this table the cosmological constant has no 
significant effect on the properties of a strange star when its value is less 
than $10^{-15}\, \mbox{cm}^{-2}$. This table generalizes our results, whereas 
in the graphs only large values of $\Lambda$ are taken to show the 
changes distinctly and also to visualize the variation pattern or dependency 
of the same with $\Lambda$ efficiently. Again some Ref.s 
\cite{Bordbar,Liu,arbanil,wei} suggest a possibility of varying $\Lambda$ 
with the cosmological evolution. Similarly, in some modified gravity theories 
like $f(R)$ gravity in Palatini formalism, the field equations can be written 
in GR form with an effective $\Lambda$ term \cite{pano}. This indicates 
the fact that the cosmological evolution obtained by using a feasible $f(R)$ 
gravity model can mimic the $\Lambda$CDM model in the present and early 
universe effectively. However, depending on the functional form and 
uniqueness of such $f(R)$ models, the effective $\Lambda$ term varies 
differently \cite{dj, santos}. Therefore, consideration of different values 
of $\Lambda$ in a wide range can be helpful to see the possible impacts of 
it on the strange star structures in theories with varying dark energy or 
$f(R)$ gravity in Palatini formalism, where the effective $\Lambda$ term is 
not a constant. However, we agree that a functional variation of $f(R)$ 
gravity models can have other impacts on the strange star properties also and 
we leave it as the future scope of our study. An important outcome of the study 
is that a $\Lambda$ value below $1\times10^{-15}\, \mbox{cm}^{-2}$ does not 
have significant observable impacts on the maximum mass, maximum radius, 
radial oscillations, and echo frequencies of a strange star with the MIT Bag 
model and linear EoSs. The boundary values of $\Lambda$ are found to be the 
same for these two EoSs. As these both EoSs are stiff EoSs, with almost similar 
compactness so the boundary values on $\Lambda$ are similar for them.

Another important aspect of the study is that the study gives a theoretical 
idea of the structure of a strange star in terms of a few properties like 
maximum mass, maximum radius, radial oscillations, and echo frequencies in 
both de Sitter and anti-de Sitter regime. The study with positive values of 
cosmological constant $\Lambda$ (which corresponds to de Sitter space) gives 
a more realistic description of strange stars. These are relevant from the 
observational point of view and also used in the context of the dark energy 
model of the universe. On the other hand, the negative cosmological constant 
$\Lambda$ corresponding to anti-de Sitter space is important for anti-de 
Sitter/conformal field theory correspondence as mentioned earlier.

To know the enigmatic stellar interior of compact stars using asteroseismology 
several ground-based missions are proposed in addition to few early space 
missions in this regard. Some of such important missions are MOST \cite{most}, 
CoRoT \cite{corot}, BRITE \cite{brite}, Kepler/K2 \cite{kepler}, 
PLATO \cite{plato}, and TESS \cite{tess} missions. The unplanned pioneer 
mission in the field of observational asteroseismology was NASA's WIRE 
mission \cite{wire}. The Canadian MOST mission was the first space telescope 
dedicated to asteroseismology and was proposed to monitor relatively bright 
stars. The French-led CoRoT mission was a great success for massive star 
variability studies. The BRITE-constellation of nanosatellites is another 
important photometry mission available for asteroseismology and it is ranked 
amongst the highest givers for providing excellent asteroseismic returns. The 
most famous of all space telescopes providing time series photometry is the 
Kepler mission launched in 2009. It is targeted to study solar like stars all 
over the HR diagram. The next generation of stellar missions such as PLATO and 
other current and future Earth based radio telescopes, such as Arecibo \cite{qui} 
or SKA \cite{ska} have the potential to put important constraints in the 
stellar properties, or to find such compact stars. Furthermore, the recently 
launched NASA's mission NICER \cite{nicer}, designed primarily to observe 
thermal X-rays emitted by several millisecond pulsars, is another ray of hope 
to detect such stars. 

Again, the current GW detectors have comparatively low sensitivity (detectable 
frequency range is $\sim$ 20 Hz - 4 kHz \cite{martynov, abbot}), so 
unfortunately the ongoing detectors does not allow the detection of such 
compact stellar oscillation frequencies. The next generation of ground-based 
GW detectors, e.g., the Einstein Telescope (ET) \cite{et} and the Cosmic 
Explorer (CE) \cite{ce} are expected to have a sensitivity much higher than 
an order of magnitude in comparison to the Advanced LIGO. These detectors are 
supposed to detect such emission and provide information on the oscillation of 
compact stars \cite{chiren}. If in near future such frequencies are detected, 
they could also provide the simultaneous measurements of compact star 
masses, tidal Love numbers \cite{benni}, frequency, damping time, amplitude of the 
modes. Furthermore, the expectations for the detection of the compact star's 
oscillations will also be increased by the launch of the eXTP \cite{extp} and 
other X-ray missions. When the detection of the radial oscillations of neutron 
stars and other compact stars will become possible, then it can be used to 
constrain the EoSs of compact stars with high precision and hence the 
corresponding stellar properties. 

From the emission of GWs, the most important modes to be detected are the 
fundamental $f$-modes and the first few pressure $p$-modes, in addition to the 
$r$-modes and $w$-modes \cite{kokkotas}. The $f$-mode of oscillation is the 
mode through which most of the gravitational radiation of the star is radiated 
away \cite{kokkotas97}. The $p$-mode is associated with the acoustic sound 
speed inside the star. These $p$- or the low order radial acoustic modes can 
be observed through the associated GW perturbations. One of the obvious sources 
of excitation of these radial (or non-radial) acoustic modes is the supernova 
explosion \cite{gpano}. The pulsating compact objects formed from such events 
emit GWs. Strong star-quakes can also arise with their associated pulsar 
glitches that can excite global stellar pulsations. Again, the compact stars 
that undergo a phase transition, e.g.\ when a neutron star collapses into a 
strange star and the coalescence of two compact stars may form a pulsating 
remnant compact star \cite{lusky}. These mechanisms of excitation generally 
give $f$-modes as well as radial $p$-modes. Moreover, due to the similar 
excitation mechanism the $f$-modes and $p$-mode can be determined using 
similar strategy \cite{glam}. Besides these excitation mechanisms, such modes 
can also be excited by tidal forces in a close eccentric binary system 
\cite{chirenti} and a resonant excitation in binaries \cite{hind}. Accretion 
in Low-Mass X-ray Binaries (LMXBs) can also excite these modes \cite{ander}. 
Star-quakes caused by cracks in the crust, magnetic reconfiguration or any 
other dynamical instabilities are also associated with the excitation of 
stellar oscillation modes \cite{franco, tsang}.

It is also worth mentioning that in actual practice, the large amplitude of 
oscillation modes only occurs in catastrophic situations, such as in core-collapse supernovae. Such events are highly non-spherical and hence in such 
cases, the non-radial modes could become more applicable. So a possible 
extension of our present work would be to study the non-radial oscillations 
of strange stars and the corresponding GW signals associated with them. In 
\cite{abedi}, the authors have claimed a tentative detection of echoes from 
the remnant of GW170817 event at a frequency $\simeq 72\,\mbox{Hz}$. Although 
the consistent mass ($2.6-2.7\,\mbox{M}_\odot$) of this remnant almost lies 
within the range of masses of strange stars predicted by our EoSs with 
different $\Lambda$ values, the GWE frequencies found from our calculations are 
much higher (in $\mbox{kHz}$ range) than this tentative detection value. The 
basic reason for the difference between this result and our results is that 
instead of considering the remnant of GW170817 as an exotic compact object 
(ECO), we have considered the star of this mass range as a strange star. 
Further, while considering the EoSs, we have neglected the possible temperature 
effect on EoS. Again we choose strange stars as non-rotating ones and solved 
the TOV equations for the static stellar model while investigating these 
properties. So, the study of rotating, anisotropic strange stars or, other 
possible ECO is a topic that will be considered for future work.

\section*{Acknowledgments}
JB would like to thank Dibrugarh University, India for the financial support 
through the grant `DURF (Extension)-2020-21' while carrying out this work.


\bibliographystyle{apsrev}
\end{document}